%% file: main.tex
\documentclass[acmsmall]{acmart}
\input{dlab_macros}
\usepackage{caption}
\usepackage{subcaption}
\usepackage{csquotes}

\renewcommand{\mkbegdispquote}[2]{\itshape}

\usepackage{wrapfig}

\begin{document}


\title[Anticipated versus Actual Effects of Platform Design Change: A Case Study of Twitter's Character Limit]{Anticipated versus Actual Effects of Platform Design Change: A Case Study of Twitter's Character Limit}

\author{Kristina Gligori\'c}
\affiliation{%
  \institution{EPFL}
  \country{Lausanne, Switzerland}
}
\email{kristina.gligoric@epfl.ch}

\author{Justyna Częstochowska}
\affiliation{%
  \institution{EPFL}
  \country{Lausanne, Switzerland}
}
\email{justyna.czestochowska@epfl.ch}

\author{Ashton Anderson}
\affiliation{%
  \institution{University of Toronto}
  \country{Toronto, Canada}
}
\email{ashton@cs.toronto.edu}

\author{Robert West}
\affiliation{%
  \institution{EPFL}
  \country{Lausanne, Switzerland}
}
\email{robert.west@epfl.ch}

\renewcommand{\shortauthors}{Gligori\'c et al.}

\begin{abstract}

The design of online platforms is both critically important and challenging, as any changes may lead to unintended consequences, and it can be hard to predict how users will react. Here we conduct a case study of a particularly important real-world platform design change: Twitter's decision to double the character limit from 140 to 280 characters to soothe users' need to ``cram'' or ``squeeze'' their tweets, informed by modeling of historical user behavior.
In our analysis, we contrast Twitter's anticipated pre-intervention predictions about user behavior with actual post-intervention user behavior: Did the platform design change lead to the intended user behavior shifts, or did a gap between anticipated and actual behavior emerge?
Did different user groups react differently?
We find that even though users do not ``cram'' as much under 280 characters as they used to under 140 characters, emergent ``cramming'' at the new limit seems to not have been taken into account when designing the platform change. Furthermore, investigating textual features, we find that, although post\hyp intervention ``crammed'' tweets are longer, their syntactic and semantic characteristics remain similar and indicative of ``squeezing''. Applying the same approach as Twitter policy-makers, we create updated counterfactual estimates and find that the character limit would need to be increased further to reduce cramming that re-emerged at the new limit.
We contribute to the rich literature studying online user behavior with an empirical study that reveals a dynamic interaction between platform design and user behavior, with immediate policy and practical implications for the design of socio\hyp technical systems. 

\end{abstract}

\setcopyright{acmlicensed}
\acmJournal{PACMHCI}
\acmYear{2022} \acmVolume{6} \acmNumber{CSCW2} \acmArticle{491} \acmMonth{11} \acmPrice{15.00}\acmDOI{10.1145/3555659}


\maketitle

\section{Introduction}
\label{sec:intro}

\input{1introduction.tex}

\section{Background and Related Work}
\label{sec:related}

\input{2related.tex}

\section{Materials and Methods}
\label{sec:methods}

\input{3methods.tex}

\section{Results}
\label{sec:results}

\input{4results.tex}

\section{Discussion}
\label{sec:discussion}

\input{5disscusion.tex}

\section{Conclusion}
\label{sec:conclusion}

\input{6conclusion}

\section*{Acknowledgments}

We are grateful to Léonore Guillain for early help with data analysis. The EPFL Data Science Lab acknowledges support from Microsoft (Swiss Joint Research Center),
Swiss National Science Foundation (grant 200021\_185043),
Collaborative Research on Science and Society (CROSS),
European Union (TAILOR, grant 952215),
Facebook,
and Google. Ashton Anderson acknowledges support from the Natural Sciences and Engineering Research Council.

\received{January 2022}
\received[revised]{April 2022}
\received[accepted]{August 2022}

\bibliographystyle{ACM-Reference-Format}
\bibliography{bibliography}

\setcounter{figure}{0}
\setcounter{table}{0}
\setcounter{section}{1}
\setcounter{subsection}{0}
\setcounter{equation}{0}
\makeatletter
\renewcommand{\thefigure}{S\arabic{figure}}
\renewcommand{\thetable}{S\arabic{table}}
\renewcommand\thesubsection{\Alph{subsection}}
\renewcommand\theequation{S\arabic{equation}}

\section*{Appendix: Supplemental Materials}
\label{sec:app}

\input{7appendix}

\end{document}
\endinput

%% file: dlab_macros.tex
\usepackage[utf8]{inputenc}
\usepackage[T1]{fontenc}
\usepackage{hyphenat}
\usepackage{xspace}
\usepackage{amsmath}
\usepackage{amsfonts}
\usepackage{hyperref}
\usepackage{url}
\usepackage{booktabs}
\usepackage{multirow}
\usepackage{makecell}
\usepackage{caption}
\usepackage{minibox}
\usepackage{bbm}
\usepackage{graphicx}
\usepackage{balance}
\usepackage{mathtools}
\usepackage{color}
\usepackage{marvosym}
\usepackage{ifthen}
\usepackage{textcomp}
\usepackage{enumitem}
\usepackage{verbatim}
\usepackage{algorithm}
\usepackage{algorithmic}
\usepackage{numprint}
\usepackage{balance}

\usepackage{amsthm}
\theoremstyle{plain}

\newcommand{\chatoDisplayMode}[1]{#1}

\definecolor{MyRed}{rgb}{0.6,0.0,0.0} 
\definecolor{MyBlack}{rgb}{0.1,0.1,0.1} 
\newcommand{\inred}[1]{{\color{MyRed}\sf\textbf{\textsc{#1}}}}
\newcommand{\frameit}[2]{
  \begin{center}
  {\color{MyRed}
  \framebox[.9\columnwidth][l]{
    \begin{minipage}{.85\columnwidth}
    \inred{#1}: {\sf\color{MyBlack}#2}
    \end{minipage}
  }\\
  }
  \end{center}
}

\newcommand{\note}[2][]{\chatoDisplayMode{\def\@tmpsig{#1}\frameit{{\Pointinghand} Note}{#2\ifx \@tmpsig \@empty \else \mbox{ --\em #1}\fi}}}
\newcommand{\todo}[2][]{\chatoDisplayMode{\def\@tmpsig{#1}\frameit{{\Writinghand} To-do}{#2\ifx \@tmpsig \@empty \else \mbox{ --\em #1}\fi}}}

\newcommand{\rev}[1]{\textcolor{black}{#1}}
\newcommand{\minor}[1]{\textcolor{black}{#1}}

\newcommand{\abbrevStyle}[1]{#1}

\newcommand{\ie}{\abbrevStyle{i.e.}\xspace}
\newcommand{\eg}{\abbrevStyle{e.g.}\xspace}
\newcommand{\cf}{\abbrevStyle{cf.}\xspace}

\newcommand{\vs}{\abbrevStyle{vs.}\xspace}

\newcommand{\Secref}[1]{Sec.~\ref{#1}}

\newcommand{\Figref}[1]{Fig.~\ref{#1}}

\newcommand{\textcite}[1]{\citeauthor{#1} \shortcite{#1}}

\newcommand{\hide}[1]{}

\makeatletter
\newcommand{\iffont}[2]{\ifthenelse{\equal{\f@family}{#1}}{#2}{}}
\makeatother

\iffont{ptm}{
  \usepackage{mathptmx}

  \DeclareSymbolFont{greek}{OML}{cmm}{m}{n}
  \DeclareMathSymbol{\alpha}{\mathalpha}{greek}{"0B}
  \DeclareMathSymbol{\beta}{\mathalpha}{greek}{"0C}
  \DeclareMathSymbol{\gamma}{\mathalpha}{greek}{"0D}
  \DeclareMathSymbol{\delta}{\mathalpha}{greek}{"0E}
  \DeclareMathSymbol{\epsilon}{\mathalpha}{greek}{"0F}
  \DeclareMathSymbol{\zeta}{\mathalpha}{greek}{"10}
  \DeclareMathSymbol{\eta}{\mathalpha}{greek}{"11}
  \DeclareMathSymbol{\theta}{\mathalpha}{greek}{"12}
  \DeclareMathSymbol{\iota}{\mathalpha}{greek}{"13}
  \DeclareMathSymbol{\kappa}{\mathalpha}{greek}{"14}
  \DeclareMathSymbol{\lambda}{\mathalpha}{greek}{"15}
  \DeclareMathSymbol{\mu}{\mathalpha}{greek}{"16}
  \DeclareMathSymbol{\nu}{\mathalpha}{greek}{"17}
  \DeclareMathSymbol{\xi}{\mathalpha}{greek}{"18}
  \DeclareMathSymbol{\pi}{\mathalpha}{greek}{"19}
  \DeclareMathSymbol{\rho}{\mathalpha}{greek}{"1A}
  \DeclareMathSymbol{\sigma}{\mathalpha}{greek}{"1B}
  \DeclareMathSymbol{\tau}{\mathalpha}{greek}{"1C}
  \DeclareMathSymbol{\upsilon}{\mathalpha}{greek}{"1D}
  \DeclareMathSymbol{\phi}{\mathalpha}{greek}{"1E}
  \DeclareMathSymbol{\chi}{\mathalpha}{greek}{"1F}
  \DeclareMathSymbol{\psi}{\mathalpha}{greek}{"20}
  \DeclareMathSymbol{\omega}{\mathalpha}{greek}{"21}
  \DeclareMathSymbol{\varepsilon}{\mathalpha}{greek}{"22}
  \DeclareMathSymbol{\vartheta}{\mathalpha}{greek}{"23}
  \DeclareMathSymbol{\varpi}{\mathalpha}{greek}{"24}
  \DeclareMathSymbol{\varrho}{\mathalpha}{greek}{"25}
  \DeclareMathSymbol{\varsigma}{\mathalpha}{greek}{"26}
  \DeclareMathSymbol{\varphi}{\mathalpha}{greek}{"27}
  \DeclareSymbolFont{otone}{OT1}{cmr}{m}{n}
  \DeclareMathSymbol{\Gamma}{\mathalpha}{otone}{0}
  \DeclareMathSymbol{\Delta}{\mathalpha}{otone}{1}
  \DeclareMathSymbol{\Theta}{\mathalpha}{otone}{2}
  \DeclareMathSymbol{\Lambda}{\mathalpha}{otone}{3}
  \DeclareMathSymbol{\Xi}{\mathalpha}{otone}{4}
  \DeclareMathSymbol{\Pi}{\mathalpha}{otone}{5}
  \DeclareMathSymbol{\Sigma}{\mathalpha}{otone}{6}
  \DeclareMathSymbol{\Upsilon}{\mathalpha}{otone}{7}
  \DeclareMathSymbol{\Phi}{\mathalpha}{otone}{8}
  \DeclareMathSymbol{\Psi}{\mathalpha}{otone}{9}
  \DeclareMathSymbol{\Omega}{\mathalpha}{otone}{10}
  \DeclareSymbolFont{syms}{OML}{cmm}{m}{it}
  \DeclareMathSymbol{\partial}{\mathord}{syms}{"40}
  \DeclareMathAlphabet{\mathbold}{OML}{cmm}{b}{it}
  \DeclareSymbolFont{largesymbols}{OMX}{cmex}{m}{n}

}

%% file: 1introduction.tex

\rev{Online platform design is, on the one hand, critically important, given that the online content can reach and affect people worldwide, and, on the other hand, very challenging, given the intrinsic entanglement between the digital environment and user behavior. The design of platforms impacts user behavior \cite{huszar2022algorithmic,malik2016identifying}, and user behavior, in turn, shapes the platforms \cite{wagner2021measuring}. It is challenging to predict how users will respond to design changes and new platform policies as any changes to large complex socio\hyp technical systems may lead to unintended consequences.}

\rev{Here we examine an instance of a particularly important real-world platform design change:}
on 7 November 2017, Twitter suddenly and unexpectedly increased the maximum allowed tweet length from 140 to 280 characters, thus altering its signature feature. According to Twitter, this change, which we henceforth refer to as ``the switch'', was introduced to give users more space to express their thoughts, as a disproportionately large fraction of tweets had been exactly 140 characters long \cite{twitter_blog2,w1}, reflecting users' need to ``cram'' or ``squeeze'' their tweets.

The design of this platform change and the choice of the new policy was informed by modeling historical user behavior \cite{t3}. When deciding how to design the platform change, Twitter engineers and policy\hyp makers made estimates and predictions based on data from before the intervention. In particular, to estimate users' need to ``cram'' their tweets, they estimated the number of tweets impacted by the policy, that is, the number of tweets that would be longer than the character limit if they could be longer. The new policy, 280 characters, was selected since it was estimated that, \minor{if that limit were enforced~\cite{t3}, a negligibly low fraction of tweets would afterwards be impacted by cramming.}



Twitter's decision to double the maximum allowed tweet length is, therefore, a remarkable example of a platform design intervention that was informed by modeling based on publicly available historical user traces. However, it is known that there are factors that render anticipating the impact of real-world interventions challenging. Forecasts by impartial experts and policy-makers often fail to predict the \rev{outcomes} of behavioral interventions \cite{milkman2021megastudies,dellavigna2018predicting}.
Users routinely violate expectations and norms, at the individual and the population level \cite{10.1145/3274301}, interventions can elicit backfire effects \cite{10.1145/3479525,swire2020searching}, and, although practically useful, predictions of online behaviors do not necessarily allow understanding of the phenomena being predicted \cite{shulman2016predictability}.

Further challenges arise given the inherent limits to the predictability of human behaviors, offline and online, at an individual and at a population level. When it comes to \emph{offline behaviors}, across several domains, there are short-term and long-term limits to predictability \cite{salganik2020measuring, song2010humanmobility,zhao2021taxi, ramso2011}. When it comes to \emph{online behaviors}, despite an unprecedented volume of information about users, content, and historical behaviors, there are limits to prediction in the complex social systems we engage with~\cite{martin2016exploring}. In studying online behaviors further challenges arise given the complex interaction between humans and the platforms we use. \rev{Human behaviors both influence and are influenced by the design and presence of technological platforms \cite{wagner2021measuring}.}

Twitter engineers and policy\hyp makers, by their own account \cite{t3}, fitted a model to historical user behavior traces and implicitly assumed that the user behavior would not change in response to the implementation of the change. Such a static, ``no-response’’ view might or might not hold. Is it necessary to account for user response, or does the actual user behavior remain faithful to what was anticipated?


Given the aforementioned challenges of modeling online human behaviors and anticipating the impact of policy changes, 
and the fact that a remarkable platform design intervention intended to impact the behavior of millions of users was made based on predictions informed by the historical user behavior modeling, 
we perform a case study of Twitter's policy change.
We ask: \emph{Did the intervention materialize as intended? Did the predictions regarding anticipated user behavior hold? How did users adapt to this platform change?} While the effectiveness of the implemented intervention can be analyzed in relation to its impact on a wide range of user behaviors, we focus on cramming---\minor{the very behavior that led Twitter to change their defining feature by doubling the character limit.} 


\rev{Reducing cramming was important for Twitter policy\hyp makers since it was hypothesized that users' need to cram the tweets to fit the limit was creating friction to post, which in turn was suspected to be linked with users abandoning their tweets and ultimately leaving the platform \cite{twitter_blog2,w1}. Beyond the importance for Twitter (one of the leading social media platforms \cite{perrin2019share}, with content posted there reaching and affecting billions of people across the globe), examining cramming behavior and the impact of the implemented intervention on cramming is broadly of importance for human--computer interaction and social media studies for three key reasons.}

\rev{First, 
when designing socio\hyp technical systems and implementing platform changes, it is important to know whether changes can be implemented based on static analyses (as performed by Twitter) or if dynamical reasoning about how users will respond is necessary. In this regard, the findings and implications from the case of Twitter could be widely applicable to researchers and practitioners designing other policy changes and other platforms that allow the production of textual content.}

\rev{Second, a deeper understanding of users' cramming behavior on Twitter is necessary, regardless of Twitter's specific policy change motivations. It is known that the policies that a platform imposes affect the audience not only through the content delivered over the platform, but also through the characteristics of the platform itself, or, in a mantra coined by Marshall McLuhan \shortcite{mcluhan1964extensions}, ``the medium is the message''. On Twitter, the imposed length limit leads to cramming, affecting the linguistic style \cite{zhou2019remaking,jin2017will} and the success of the tweets, measured through the received engagement \cite{r9,w1,10.1145/3359147,wang2020does,wasike2013framing,wagner2017framing,schultz2017proposing,gligoric2021linguistic}. Therefore, cramming has implications for the nature and quality of discourse on the platform and the degree to which messages are likely to spread \cite{w1}. Nobody in whose research Twitter plays a role should ignore this platform change; all researchers should consider how the new length limit impacts users' cramming behavior, the content, and Twitter as a platform. Since cramming is associated with the linguistic features of the message and its success, when analyzing or modeling the textual content posted on Twitter, researchers need to know whether, after the switch, the retrieved messages are still ``compressed'' by the users or not.}

\rev{Third, understanding cramming behavior is consequential beyond Twitter. Cramming is inherent in writing text under any length constraint, online and offline. As such, cramming should be understood when studying any textual content produced under a character limit, given that the limit policy is bound to shape the content \cite{w1}. Insights about cramming behavior and its variation across subpopulations of users depending on their languages and devices are relevant to researchers studying any form of textual content produced under a length limit.}

\subsection{Research questions}

In a case study of Twitter's policy change, studying an instance of platform design change, we aim to shed light on how this change has impacted user behavior. Now that the length limit has been changed and it is possible to tweet longer than 140 characters, we can compare estimates made before the intervention with how user behavior has actually evolved. We aim to fill the gaps in the existing literature through the following core research question: 

\begin{quote} 
    \centering 
    \textbf{RQ:} Did the introduction of the 280-character limit lead to the intended user behavior shifts, or did a gap between anticipated and actual behavior emerge?
\end{quote}

\noindent In particular, studying user modeling that informed design change on a major platform, we investigate the following dimensions of the above question:
\begin{enumerate}
\item[\textbf{RQ1}] \textit{Gaps between predictions and emerging user behavior:} Did the platform design intervention address the problem of cramming as intended?
\item[\textbf{RQ2}] \textit{Cramming across languages:} How are different user populations affected by the same global platform design change?
\item[\textbf{RQ3}] \textit{Fluidity of counterfactual estimates:} \rev{How do the predicted policy effects diverge after user behavior shifts?} 


\end{enumerate}

\subsection{\rev{Contributions}}

\rev{In this work, we report a novel empirical study of a design decision that shaped a socio\hyp technical system and impacted millions of users. Our results reveal a dynamic interaction between platform design and user behavior, with immediate policy and practical implications for the design of socio\hyp technical systems. We call for the development of cautious approaches that aim to consider multiple factors when designing a platform change, including the dynamic nature of user response and the impact the change will have on different populations depending on users' languages and devices. }

\subsection{Summary of main findings} 

Using a 1\% sample of all tweets spanning the period from 1 January 2017 to 31 October 2019, we model tweet length over time (illustrated in \Figref{diagram}). We find that gaps between anticipated and actual user behavior emerged after the intervention (RQ1). Initially, the estimated amount of cramming at 140 characters was aligned with actual user behavior (\Figref{timeseries}). However, actual user behavior eventually diverged from anticipated user behavior. While cramming at 140 characters sharply decreased after the introduction of 280 characters, cramming, although less drastic, shifted to the new length limit. Furthermore, examining tweet text, syntactic (\Figref{pos_tags} and~\ref{pronouns}) and semantic (\Figref{fig:politics}) indicators also provide evidence of cramming that emerged at the new length limit. Overall, these gaps between anticipated and actual user behavior are more pronounced on the Web interface compared to mobile devices. 

Studying how different user populations are affected by the global platform design change (RQ2), we find that \minor{cramming in a language at 140 characters before the switch is correlated with cramming in that language at 280 characters after the switch} (\Figref{figlangs2}), indicating that Twitter is used differently across languages, or that some languages might need more or fewer characters to express the same amount of information. 

Finally, we consider hypothetical interventions that would reduce the cramming that emerged post\hyp intervention (RQ3). We find that as user behavior shifts in response to a platform change, the estimated effects of hypothetical policies change as well (\Figref{runover1}). Given that 280 characters were selected to make a vast majority of tweets fit the limit, post\hyp intervention, since the cramming re\hyp emerged, the necessary number of characters would have to be increased further to achieve the same objective (\Figref{runover2}).

\subsection{Implications} 

Our case study has immediate implications for platform design and future platform changes. Our results emphasize a dynamic interaction between platform design and user behaviors. The length limit was doubled in order to reduce users' need to cram their tweets. However, the intervention did not entirely solve the issue, as cramming re-emerged at the new length limit (\Figref{timeseries}). \minor{Before the intervention, the modeled data was ``collected under the policy'', with character limit shaping the data that Twitter subsequently based their measurements on. When modeling in such a static regime, the validity of estimates is threatened, and it is complicated to evaluate alternative policies before their deployment. As the new policy is deployed, new behaviors and new data are collected under a different policy which elicits behaviors that are not anticipated---although can be explained after the fact. This fusion of platform design decisions and user behaviors can lead to feedback loops and calls for more cautious approaches that take into account the dynamic nature of user response. These findings highlight the fluidity of online behaviors and have direct implications for large\hyp scale user behavior studies, human--computer interaction, and platform design.}




%% file: 2related.tex
\subsection{Previous work studying Twitter's character limit change} 

\minor{Early existing work that studied Twitter users' attitudes toward the new 280-character limit \cite{10.1145/3293339.3293349} discovered varying initial reactions ranging from anticipation, surprise, and joy to anger, disappointment, and sadness. Early studies also revealed a low initial prevalence of long tweets immediately following the switch \cite{twitter280blog} and studied the short-term impact that the length limit had on linguistic features and engagement \cite{w1,boot2019character}, finding that in response to a length constraint, users write more tersely, use more abbreviations and contracted forms, and use fewer definite articles.}

\minor{The impact of the switch was also studied in the specific context of political tweets \cite{jaidka2019brevity}, showing that doubling the length limit led to less uncivil, more polite, and more constructive discussions online. Whereas these early studies necessarily had to consider short-term effects, much less is known about the long-term effects of the switch on user behavior and about {whether Twitter's intention was achieved or not after the implemented intervention}. The present paper constitutes the first attempt to bridge this gap with a long-term study spanning several years.}

\subsection{Counterfactual policies and user responses}\label{sec:relfeedback}

Platform design decisions are often shaped by data\hyp driven user studies \cite{ahn2007impact,cusumano2010technology,belavina2020rethinking}. Broadly, one can think about data\hyp driven user studies by considering the degree to which the study is \emph{explanatory} (\ie, focusing on identifying and estimating causal factors of user behavior), and the degree to which the study is \emph{predictive} (\ie, focused on predicting user behavior). \citeauthor{hofman2021integrating}~\cite{hofman2021integrating} propose integrating predictive and explanatory modeling, for example, by analyzing proposed counterfactual policies and interventions, quantifying impacts on specific behaviors in the short and long run, ahead of their implementation.

Machine learning models are often employed for both predictive and explanatory purposes. However, existing machine learning models typically rely on the assumption that the data, after deployment, resembles the data the model was fit on. As machine learning models are increasingly used to make consequential decisions and users react to the deployed models, this assumption is violated. The reactions create challenges to the deployment of machine learning algorithms in the real world. Studying such reactions, referred to as ``strategic feedback'', has enabled new avenues of machine learning research \cite{hardt2016strategic,miller2020strategic} that takes into account the feedback of the environment. Behavioral economics aims to understand and model people's strategic responses as well \cite{morecroft2015strategic}. Twitter's effort to design a real-world platform intervention based on estimates and predictions derived from historical user behavior represents an example of integrative data\hyp driven study that can be potentially be impacted by the users' reactions.

\rev{However, despite the advances in understanding and incorporating the feedback of the environment, in practice, a static view is often assumed. Twitter engineers and policy-makers modeled historical user behavior traces, apparently assuming that user behavior would not change in response to the intervention. It is unknown whether assuming such a static view is justified or not. We aim to fill this gap by answering the question of whether it is necessary to account for user responses or not (\cf RQ1). Our analyses of whether the intervention led to intended user behavior shifts will have implications for the future real\hyp world development of integrative user modeling approaches.}

\subsection{Twitter communication and supporting features: studies of use and conventions}\label{sec:relcomm}

Previous work has extensively studied communication taking place on Twitter and the specific features that support it, most importantly: retweets \cite{boyd2010tweet}, hashtags \cite{wikstrom2014srynotfunny,page2012linguistics}, quotes \cite{garimella2016quote}, and emojis \cite{pavalanathan2016more}. Previous work has also investigated linguistic conventions on Twitter, the patterns of their emergence \cite{kooti2012emergence,kooti2012predicting}, how users align to them in conversations \cite{doyle2016robust}, how they diffuse \cite{centola2018experimental,chang2010new}, and how they continuously evolve \cite{cunha2011analyzing}. Additionally, previous work has studied how patterns of adoption of supporting features, as well as the linguistic style used on the platform more broadly, varies across numerous dimensions \cite{shapp2014variation}, including gender \cite{ciot2013gender}, political leaning \cite{sylwester2015twitter}, age and income \cite{flekova2016exploring}; but also within accounts 
\cite{clarke2019stylistic}.

\rev{Still, little is known about how users adapted to the new character limit feature in the long run, as opposed to in the short term. How did the different subpopulations of users change their communication patterns in response to the intervention (\cf RQ2)? We contribute to the literature in usage and conventions by studying how brevity conventions and norms evolve after the introduction of the new character limit feature.}


\subsection{Message framing and its linguistic features}

Focusing specifically on the linguistic feature of the character length, previous work has studied how the imposed length constraint on Twitter and other microblogging platforms affects the dialogues and the linguistic style \cite{zhou2019remaking,jin2017will}, and the success of the message measured through the received engagement \cite{r9,w1,10.1145/3359147,wang2020does,wasike2013framing,wagner2017framing,schultz2017proposing,gligoric2021linguistic}. More broadly, HCI, philology, communication, marketing, education, and psychology scholars have investigated conciseness and its benefits in many different contexts \cite{laib1990conciseness,vardi2000brevity,sloane2003say}. The length limit can be thought of as a constraint that determines the format of the content that can be produced. Such constraints are of interest to scholars since they are often thought to have a positive impact on the quality of produced creative content \cite{i1,i2,i3,i4,i5,i8}. 

Beyond length, numerous previous studies have investigated the question of what wording makes messages successful in online social media, often formulated as the task of predicting what makes textual content become popular \cite{berger2012makes,guerini2011exploring,lamprinidis2018predicting}. In the specific case of Twitter, in addition to characterizing how language is used on the platform in general \cite{murthy2012towards,levinson2011long,hu2013dude,eisenstein2013bad}, researchers have investigated the correlation of linguistic signals and user's features with the propagation of tweets \cite{artzi2012predicting,bakshy2011everyone,r9,doi:10.1080/15534510.2016.1265582,gligoric2020experts}. 

\rev{Yet, little is known about the long-term impact of the implemented intervention on the linguistic features of tweets. How did users modify their messages in adaptation to the new character limit feature, and are there still indicators of cramming (\cf RQ1)? How many characters are needed to reduce the cramming, if any, after the intervention (\cf RQ3)? We aim to provide new perspectives on how the new length limit impacts the linguistic characteristics of the messages after the intervention, in the long run. Given this active area of research, it remains important to understand how the shifts in platform design impact the language on the platform.}

%% file: 3methods.tex
\subsection{Data}

We base our user modeling on the publicly available 1\% sample of tweets, spanning the period between 1 January 2017 and 31 October 2019, available on the Internet Archive.%
\footnote{\url{https://archive.org/details/twitterstream}} We consider only original tweets (\ie, we discard retweets).%
\footnote{Our main analyses study tweets in isolation. However, Twitter users have long been working around the character limit by splitting a long piece of text into a sequence of length-compliant tweets. We consider threaded tweets in a supplementary analysis (Appendix A).}
We study the 23 biggest languages: three languages where the switch did not happen (Japanese, Korean, and Chinese), and 20 where it happened, each language with more than 2M tweets in total. The switch did not happen in Japanese, Korean, and Chinese because the 140-character limit was not as restrictive there as in other languages, since more information can be conveyed with the same number of characters \cite{twitter_blog1,twitter_blog2}. \rev{We focus on the most common sources of tweets: the Web interface and mobile applications, as specified by the \textit{source} field present in the collected tweet objects. We keep tweets posted by Twitter applications for iPhone, Android, iPad, Windows Phone, and their Lite versions and mobile Web clients and desktop Web clients. We disregard tweets posted by all other unofficial and automated sources and third-party applications as a proxy for bots (9.5\% of original tweets are discarded on average across studied languages).}

\begin{figure}[t]
     \begin{minipage}[b]{0.4\textwidth}
     \centering
     \hspace{0.3cm}
     \includegraphics[width=0.63\textwidth]{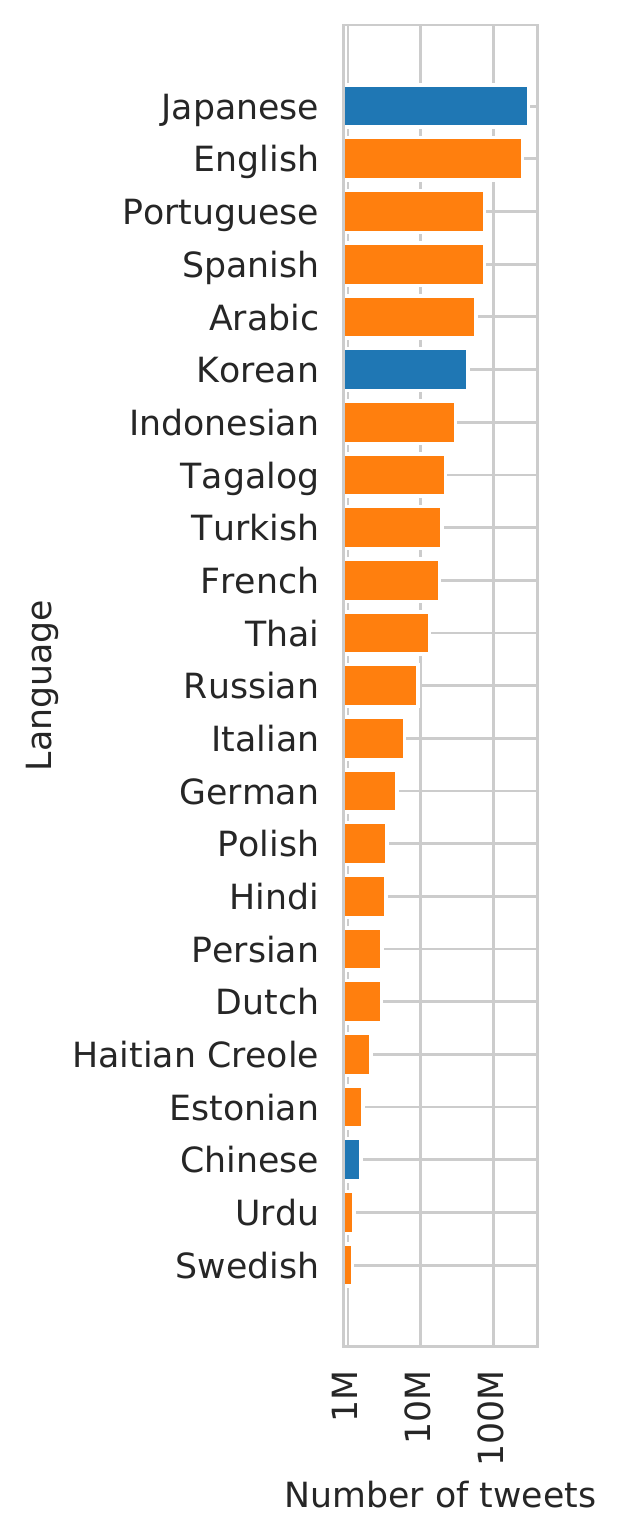}
     \subcaption{}
     \label{fig:counts}
     \end{minipage}
     \hfill
     \begin{minipage}[b]{.59\textwidth}
     \centering
     \hspace{0.2cm}
     \includegraphics[width=0.78\textwidth]{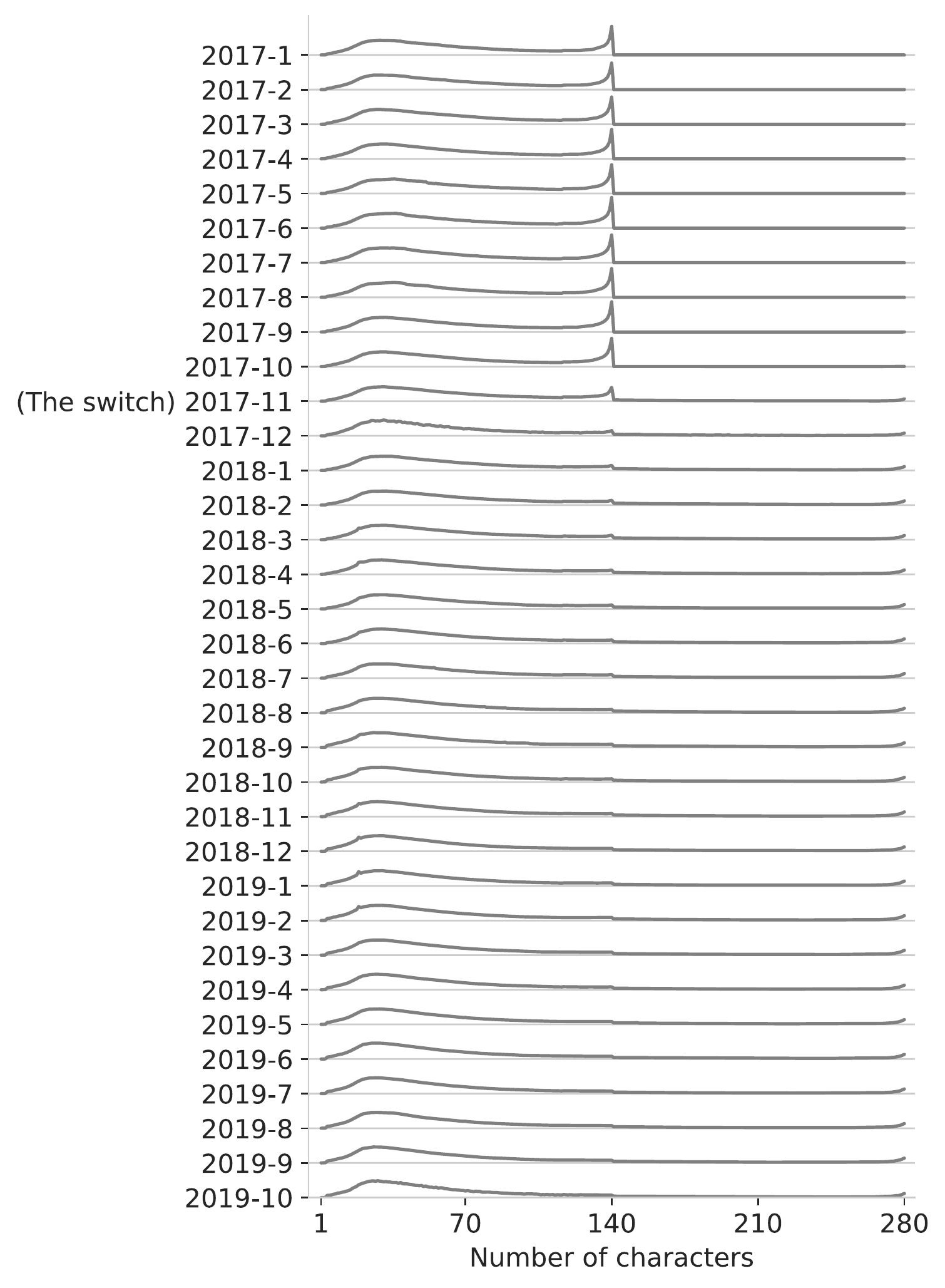}\hfill{}
     \subcaption{}
     \label{fig:hists}
     \end{minipage}
 \caption{\textbf{Twitter dataset statistics.}
(a) Number of original tweets in the 1\% sample posted between 1 January
2017
and 31 October~2019, across the 23 studied languages where the switch happened (\textit{orange}) and did not happen (\textit{blue}).
(b)~Normalized monthly tweet length histograms.
 }
 \label{descriptive}
\end{figure}

With the above, there are between 1M and 1.5M daily original tweets. In \Figref{fig:counts} we show the exact number in total across languages. We note that the posts are sampled at the community level. We stay at describing the community-level behaviors as opposed to user-level behaviors since user-level information is incomplete (in expectation, we have 1\% of tweets posted by a fixed user).

\subsection{Tweet length: counting characters}

The focus of our study is the character limit, its change, and the impact on user behaviors, captured via the length of the posted tweets. To that end, we carefully count the number of characters based on the official documentation.%
\footnote{\url{https://developer.twitter.com/en/docs/counting-characters}} Tweet length is counted using the Unicode normalization of the tweet text. The tweet text is selected from the tweet object using the \textit{displayed text range} field, discarding any characters not counted towards the length limit. 

The text content of a tweet could contain up to 140 characters (or Unicode glyphs) before the switch, and 280 after the switch. An emoji sequence using multiple combining glyphs counts as multiple characters. Chinese, Japanese, or Korean glyphs were counted as one character before, and as two characters after, the introduction of the 280-character limit. Therefore, a tweet composed of only Chinese, Japanese, or Korean text can only have a maximum of 140 of these types of glyphs after the switch.

\begin{figure}[t]
    \begin{minipage}{\textwidth}
        \centering
        \includegraphics[width=0.99\textwidth]{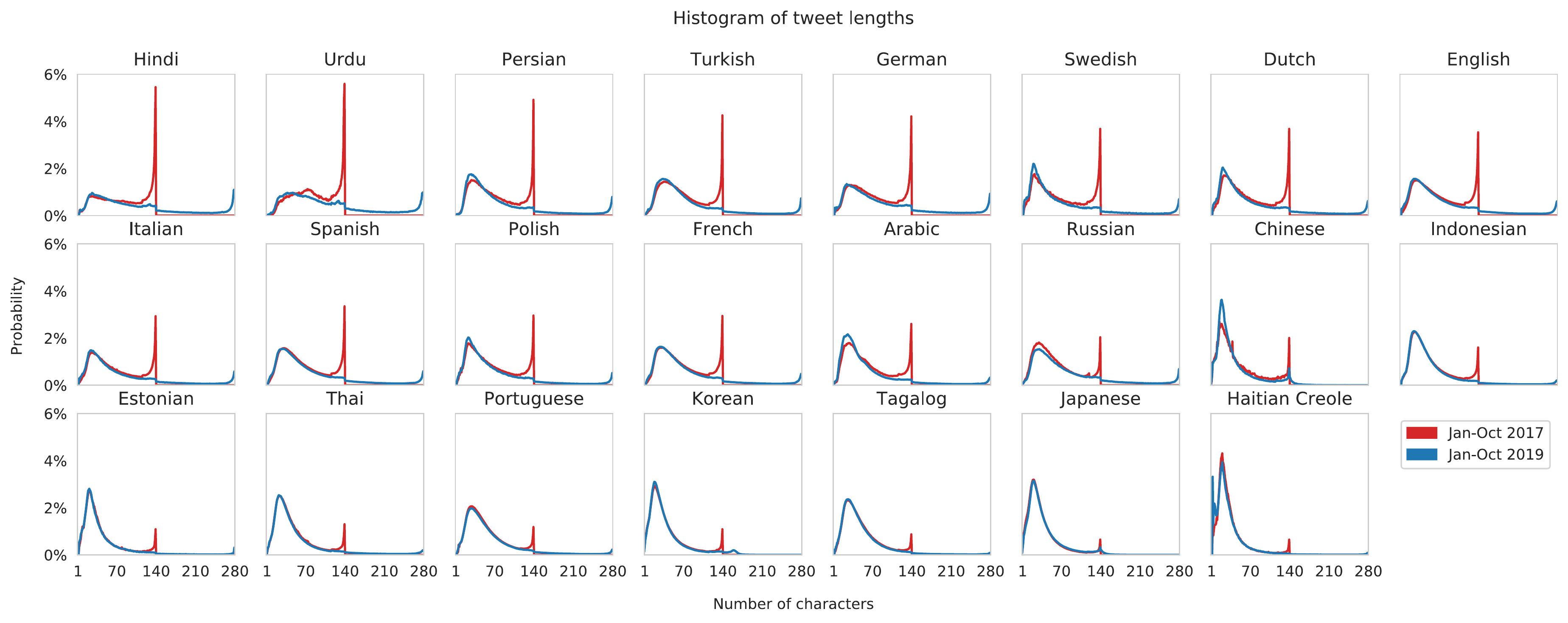}
    \end{minipage}
    \caption{\textbf{Tweet-length distributions for 23 languages,} for the periods before (\textit{red}) and after (\textit{blue}) the 280\hyp character limit was introduced. Languages sorted by prevalence of 140 characters before the switch.}
    \label{fig:histlangs}
\end{figure}

We illustrate the prevalence of different tweet lengths over time (the fraction of tweets of a given number of characters) in \Figref{fig:hists}, and for individual languages in \Figref{fig:histlangs}. We note that the first peak in the tweet length distribution, consistently between 25 and 30 characters, remains unchanged (\ie, the mode of the distribution is stable). Before the doubling of the character limit, we observe a drastic peak in the prevalence of 140-character tweets, reflecting users' need to cram their tweets \cite{twitter_blog2,w1}. After the character limit doubling, we observe a sharp decline of 140-character tweets and an increase in 280-character tweets. Similar to the overall view, across languages the first peak and the mode of the distribution are constant, and the interesting character length ranges are near 140 and near 280 characters, where tweets are impacted by cramming.

\subsection{Tweet length: modeling the impact of the character limit}

\subsubsection{Background} 

Before implementing the character length change, Twitter engineers and policy\hyp makers aimed to answer the following questions~\cite{t3}: \blockquote{\textit{Are 140 characters an adequate limit for all languages? Do people Tweeting in Japanese have too much space? Do those Tweeting in English not have enough? How often were people trying to craft Tweets which ended up over 140 characters? And by how much?}} Before the intervention, these questions seemed impossible to answer since there were no Tweets in Twitter's database over 140 characters long. 
However, Twitter policy\hyp makers analyzed historical tweeting behavior to find answers~\cite{t3}.

\begin{figure}
    \centering
    \includegraphics[width= 0.85\textwidth]{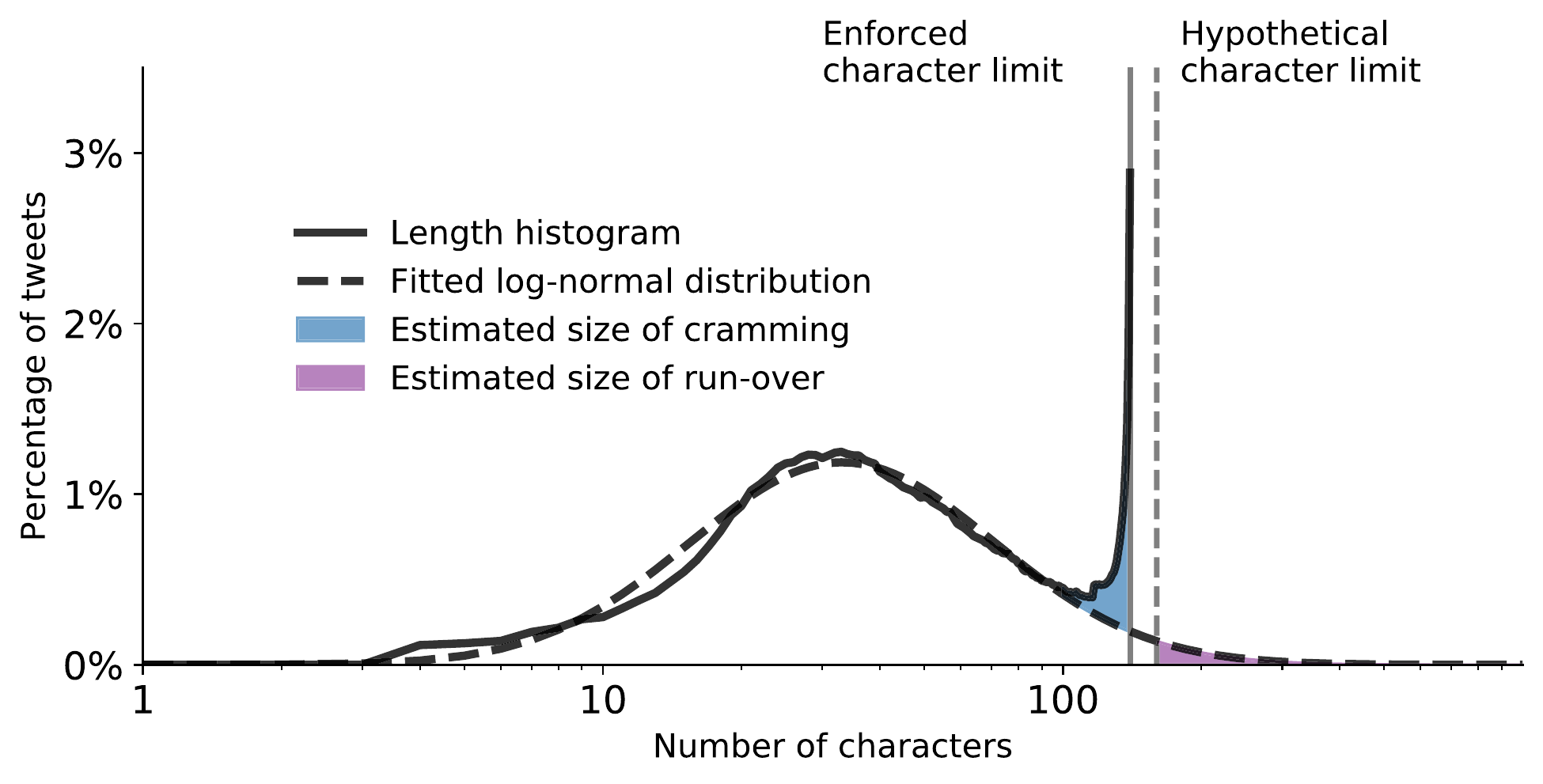}
    \caption{\textbf{Illustration of model for measuring impact of character limit.} The example depicts the length distribution of English tweets across mobile and Web devices, before the 280-character limit was introduced. The solid black line shows the length histogram; the dashed black line marks the fitted log\hyp normal distribution (logarithmic $x$\hyp axis). The solid gray vertical line marks the enforced character limit (140 characters in the example). The dashed gray vertical line marks a hypothetical character limit (150 characters in the example). \textit{Cramming} (marked as the blue area) is the deviation of the empirical distribution from the underlying log\hyp normal distribution near the character limit. \textit{Run\hyp over} (marked as the purple area) is the part of the fitted log-normal distribution that falls beyond the given number of characters. Cramming can only be used to measure people's attempts to ``squeeze'' their tweets within the \emph{enforced character limit} (solid vertical line). On the contrary, run\hyp over is used to evaluate the impact of \emph{any hypothetical character limit} (dashed vertical line).}
    \label{diagram}
\end{figure}

First, Twitter's analyses revealed that when people write text within a certain length constraint, the resulting text lengths follow a log\hyp normal distribution.

Second, it was discovered that the empirical distribution of tweet lengths deviates from the log-normal distribution near the character limit, since users try to ``squeeze'' their messages. To quantify the impact of the enforced character limit and other hypothetical character limits on users' behavior, two quantities were introduced: \emph{size of cramming} and \emph{size of run\hyp over}. The estimated size of run\hyp over is a crucial quantity, since it is used to evaluate hypothetical policies, as put by Twitter: \blockquote{\textit{9\% of English Tweets, and 0.4\% of Japanese Tweets, do not make it under 140 characters. And we will need 274 characters to make 99\% of English Tweets viable as-is}}.

Third, it was assumed that when a tweet ends up being more than 140 characters long, with probability $p$, a user deletes some characters proportional to the number of characters exceeded. With probability $1-p$, the user abandons the tweet. Then, with probability $q$, the truncated tweet constitutes a valid sentence. The user sends the tweet if it is under 140 characters at this point. Otherwise, the process is repeated. Finally, with these factors were put together, Twitter developed a model that accurately approximates the tweet length distribution with cramming~\cite{t3}. Twitter concluded that more characters are necessary to fit all tweets and that increasing the length limit might increase the fraction of users who post tweets, or, as stated by Twitter~\cite{t3}: \blockquote{\textit{It did seem reasonable to expect that people will Tweet more if there is less friction to Tweet. This looked like something worth trying.}}

In the present work, our goal is to study the above user modeling that informed Twitter's design change. To that end, we carefully implement the estimation of the size of cramming and the size of run\hyp over based on the official public documentation outlining the modeling approach \cite{t3}. We ensure that Twitter's reported pre\hyp intervention estimates in the case of English and Japanese tweets are reproduced. Then, we monitor the daily temporal evolution of these quantities, across devices and across languages. In what follows, we explain in more detail how these quantities are calculated and what they capture.

\subsubsection{Estimating the impact of the enforced character limit: the size of cramming} 

The enforced character limit impacts users' behavior, pushing people to ``squeeze'' their tweets within the allowed number of characters \cite{w1}. Due to such ``squeezing'', the empirical distribution of tweet lengths deviates from the log-normal distribution near the character limit.
Cramming at a given number of characters (illustrated in \Figref{diagram}, blue area) is, therefore, the deviation of the actual length distribution from the theoretical log\hyp normal distribution near the character limit. The underlying log-normal distribution is found by fitting a curve to the empirical tweet-length density excluding the tail emerging from cramming, using the least-squares objective. The tail exclusion is done at a point referred to as cramming threshold. The cramming threshold is found heuristically by selecting the rightmost local minimum of the tweet length density curve.

To summarize, cramming measures people's attempts to ``squeeze'' their tweets within {the enforced character limit}. Cramming is interpreted as the fraction of tweets impacted by the enforced character limit. Since the character limit was increased in order to reduce cramming \cite{twitter_blog1}, now, after the character limit was increased, we monitor the estimated size of cramming to understand the consequences of the implemented intervention.

\subsubsection{Estimating the impact of hypothetical character limits: the size of run\hyp over}

Cramming is used to measure the impact of the enforced character limit. However, cramming cannot readily be used to measure the impact of hypothetical limits surpassing the currently enforced limit, since the value of the length distribution at hypothetical unsupported tweet lengths is zero.
Run\hyp over at a given number of characters (illustrated in \Figref{diagram}, purple area) is, therefore, defined as the part of the fitted log-normal distribution that falls beyond a given number of characters. Run-over captures the fraction of tweets that would have more characters than the hypothetical length limit if it were possible.

To summarize, run\hyp over captures the fraction of tweets that would be impacted by \emph{any hypothetical character limit}. Ahead of the switch, run\hyp over was used by Twitter engineers and policy\hyp makers to evaluate the impact of counterfactual policies and to select the implemented policy (280 characters). Now, after the policy was changed, we monitor the estimated size of run\hyp over to explore the evolution of estimates and the impact of hypothetical character limits.

%% file: 4results.tex
\begin{figure}[t]
    \centering
    \includegraphics[width= 0.9\textwidth]{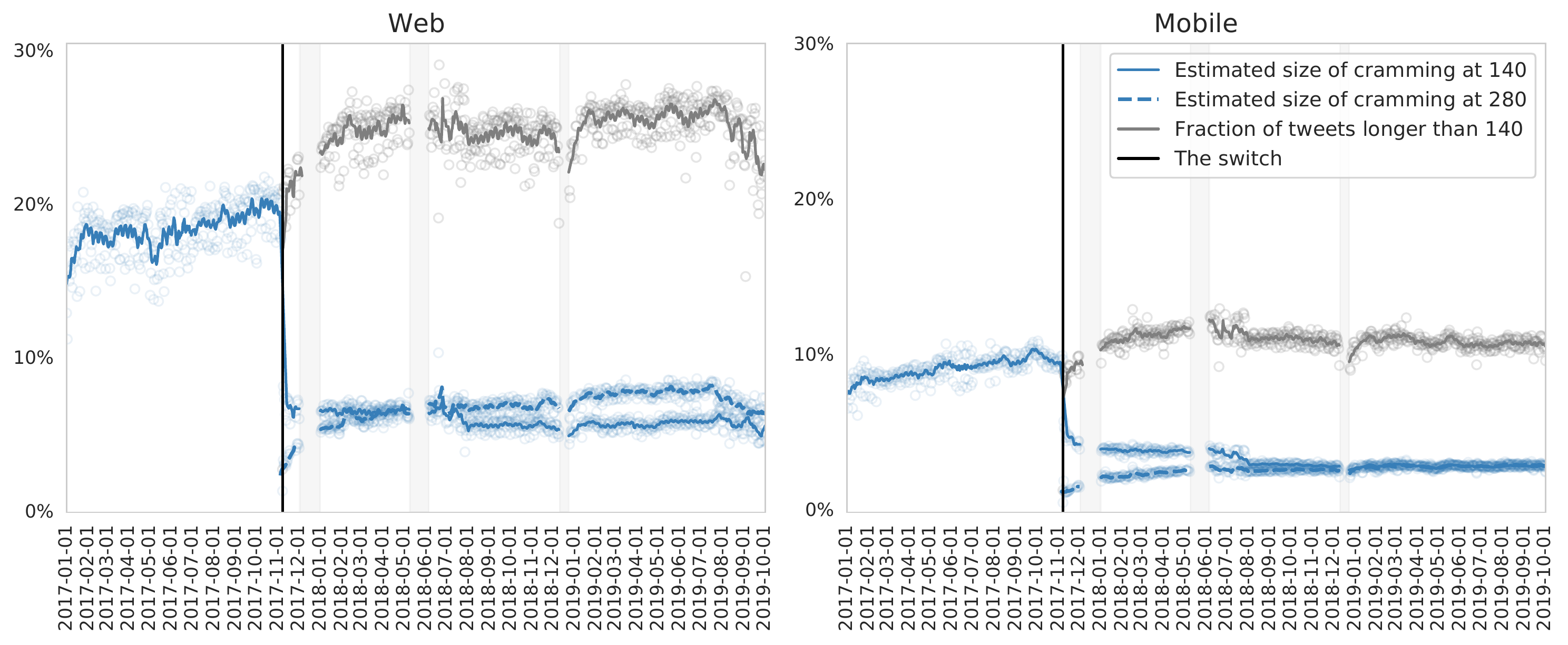}
    \caption{\textbf{Estimated \vs\ actual fraction of tweets impacted by the length limit.} In blue, the estimated size of cramming, \ie, the estimated fraction of tweets impacted by the length limit, and, in gray, the fraction of tweets longer than 140 characters. Daily quantities are indicated with a circle, and the line marks a 10-day rolling average. The quantities are shown separately for the Web interface (left) and mobile applications (right). The vertical line marks the switch and gray bands mark days with missing data. Immediately after the switch, the fraction of tweets longer than 140 characters (solid gray line) closely mirrored the estimated size of cramming (solid blue line). As time passed, usage of long tweets diverged from estimates made before the intervention. After the switch, cramming at 140 characters drastically decreased (solid blue line), while cramming at 280 characters emerged (dashed blue line).}
    \label{timeseries}
\end{figure}

\subsection{RQ1: Gaps between predictions and emerging user behavior}

Recall our guiding research question: Are there gaps between predicted and emerging user behavior? Concretely, did the platform design intervention address the problem of cramming as intended? Having access only to tweets tweeted before the switch, the fraction of tweets impacted by the limit (\ie, cramming) at 140 characters was estimated. Now, after the switch, we can compare the estimated size of cramming at 140 characters before the switch with the actual fraction of tweets that are longer than 140 characters after the switch. Contrasting these two quantities allows us to measure the gaps between estimates made before the intervention and the actual user behavior after the intervention.

To do so, we fit the model of tweet lengths (\Secref{sec:methods}) for each day of the studied period (between 1 January 2017 and 31 October 2019) across all tweets in the 20 studied languages where the 28\hyp character limit was introduced.\footnote{We take advantage of the fact that the new character limit was not introduced in Chinese, Japanese, and Korean to perform an estimation of the effect of the switch on tweet lengths that accounts for possible global platform\hyp wide changes that are not associated with the doubling of the character limit intervention (Appendix B).} We then monitor how the cramming size estimate evolves over time (\Figref{timeseries}). We consider the Web interface and mobile devices separately (11\% of tweets are sent from the Web interface \vs 89\% from mobile devices).

\subsubsection{Web interface}

Focusing on the Web interface (\Figref{timeseries}, left), before the switch, the estimated fraction of tweets impacted by the length limit, \ie, the estimated size of cramming at 140 characters, was, on average across days, 18.35\% (95\% bootstrapped CI [18.15\%, 18.55\%]). After the switch, the actual fraction of tweets longer than the previously imposed limit was, on average, 24.81\% [24.66\%, 24.95\%].

Although, in the first weeks after the 280 character limit was introduced, the actual fraction closely mirrored the prior estimates, as time passed, users' tweeting behavior drifted away from the cramming estimates that had informed the policy change, and stabilized after around six months.

Furthermore, cramming at 140 characters drastically decreased after the switch, while cramming at 280 characters emerged. Before the switch, the estimated size of cramming at 140 characters was, as stated above, 18.35\% on average across days. After the switch, the estimated size of cramming at 280 characters is 6.88\% [6.8\%, 6.96\%]. While cramming at 140 characters sharply decreased after the introduction of 280 characters, cramming, although less drastic, shifted to the new length limit. Cramming at the new length limit slowly increased and also stabilized after around six months after the intervention.

The introduction of 280 characters reduced cramming at the respective limit by 62.5\% (from 18.35\% to 6.88\%). However, a substantial fraction of tweets (around 1 in 15) is still impacted by the 280\hyp character limit on the Web interface. By all we know, the emergence of cramming at the new length limit was not anticipated in the modeling approach that informed the platform change (\Secref{sec:methods}).

\begin{figure}
    \begin{minipage}[b]{0.89\textwidth}
    \centering
    \includegraphics[width=\textwidth]{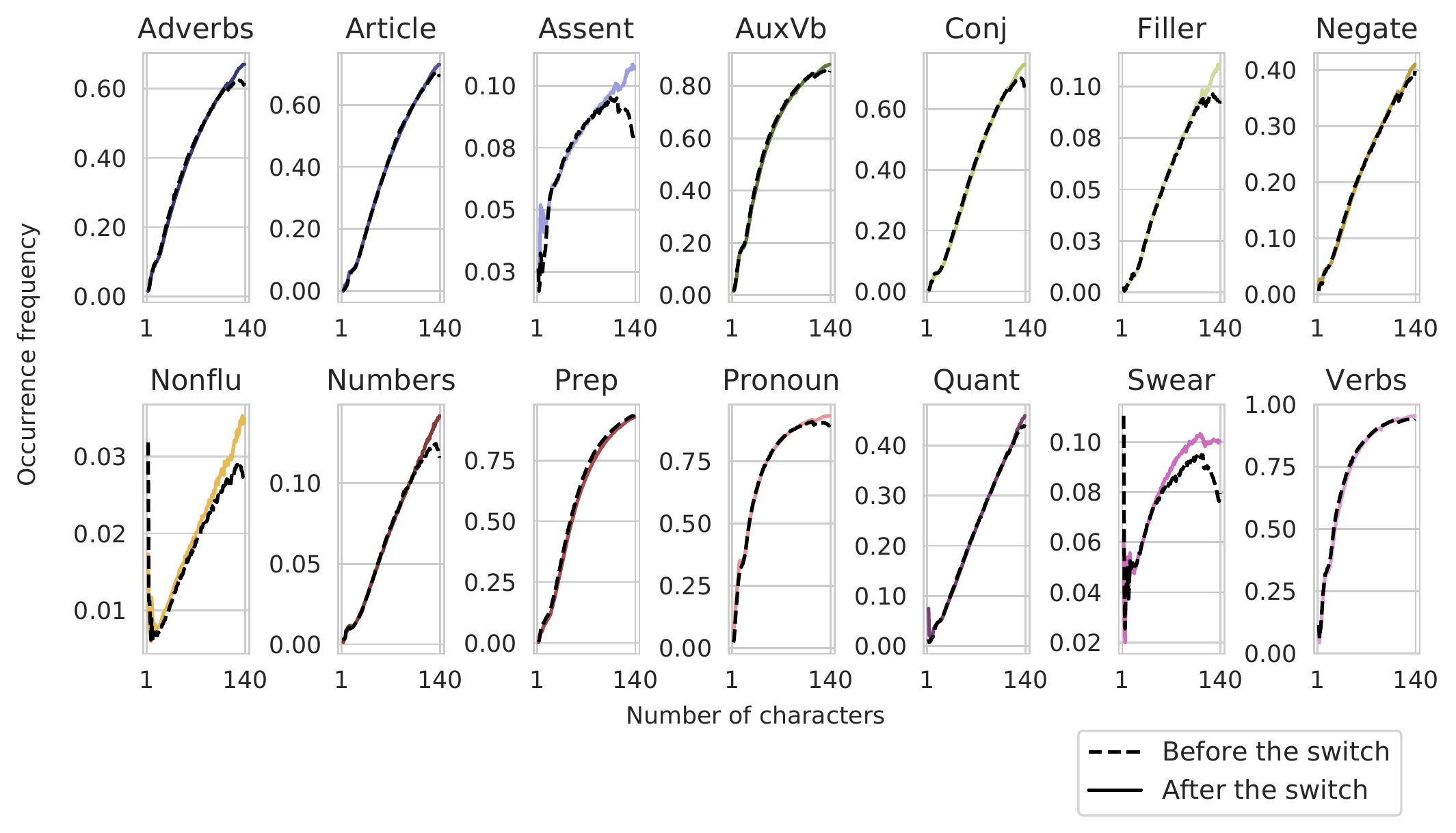}
    \subcaption{Before \vs\ after the switch}
    \label{fig:10}
    \end{minipage}
    \hfill
    \begin{minipage}[b]{0.89\textwidth}
    \centering
    \includegraphics[width = \textwidth]{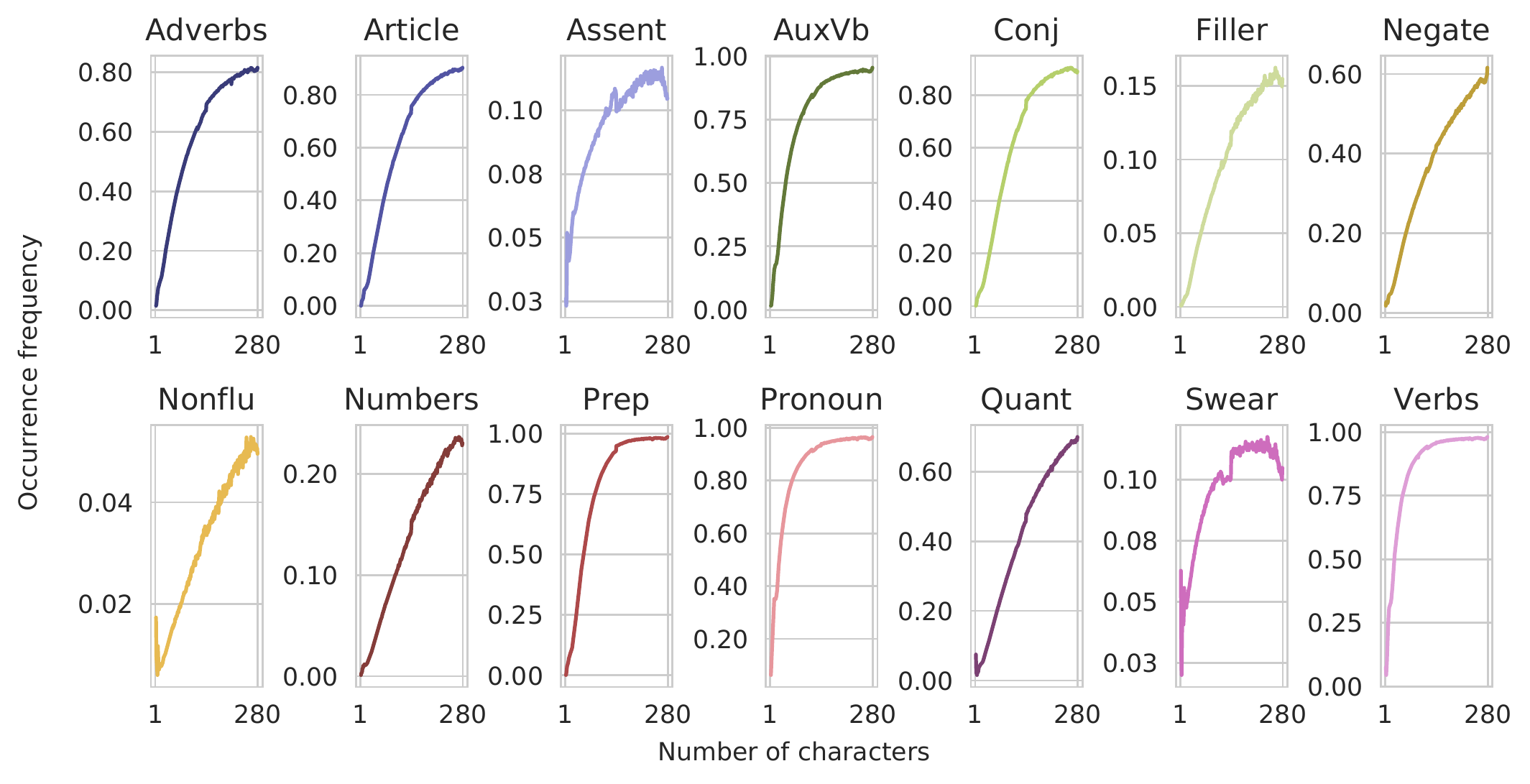}
    \subcaption{After the switch}
    \label{fig:11}
    \end{minipage}
\caption{\textbf{Syntactic indicators of cramming: part-of-speech (POS) tag frequency across tweet lengths.} (a)~Occurrence frequency of POS tags across tweets of different character lengths in the period before the switch \vs\ after the switch (1--140 characters). The dashed black line represents this quantity across tweets posted in the period before the switch (\ie, under the 140\hyp character limit), and the solid colored line in the period after the switch (\ie, under the 280\hyp character limit). \rev{The largest gap is observed for swear words and spoken categories (nonfluencies, fillers, and assent), adverbs, and conjunctions, all nonessential parts of speech that are frequently deleted in the process of ``squeezing'' a message to fit a length limit. No gap is observed for verbs and negations, essential parts of speech.}
(b)~Occurrence frequency of POS tags across tweets posted in the period after the switch (\ie, under the 280\hyp character limit) of different character lengths (1--280 characters). \rev{Among the 280\hyp character tweets after the switch, we observe patterns similar to those associated with 140\hyp character tweets before the switch, \eg, a dip in the frequency of spoken categories, conjunctions, and numbers---traces typical of cramming or ``optimizing'' a message to fit a length limit.} POS tags are sorted alphabetically. Note the different $x$- and $y$-axis scales.
\unboldmath
}
\label{pos_tags}
\end{figure}

\subsubsection{Mobile devices}

When it comes to mobile devices (\Figref{timeseries}, right), we observe that actual user behavior diverged from the cramming estimates made before the switch less drastically than in the case of the Web interface (see above). Similarly, although some cramming emerged at the new limit, it is less pronounced compared to cramming on the Web interface. The estimated size of cramming at 140 characters before the switch was 9.06\% [8.97\%, 9.15\%], whereas after the switch, 10.88\% [10.82\%, 10.93\%] of tweets were longer than the newly imposed limit. 

Compared to the Web interface, there was 50\% less cramming on the mobile devices before the switch. Similarly, little cramming emerged at the new limit. Only 2.55\% [2.52\%, 2.58\%] (or 1 in 40) tweets are impacted by the 280 character limit on mobile devices.
That is, cramming is a smaller problem on mobile devices in general. Whereas on the Web interface cramming was underestimated, and more cramming emerged at the new limit, mobile devices paint a different picture. There, user behavior remained closer to estimates made before the intervention, and little cramming emerged at the new limit.

To summarize, the modeling approach underestimated the
fraction of tweets that would be longer than the 140-character limit.
Also, cramming now emerging at 280 was apparently not considered in the modeling approach performed before the switch. Cramming was a smaller issue on mobile devices before the intervention, compared to the Web interface, and remained less problematic after the intervention.

\begin{figure}
    \begin{minipage}[t]{0.95\textwidth}
    \centering
    \includegraphics[width=0.9\textwidth]{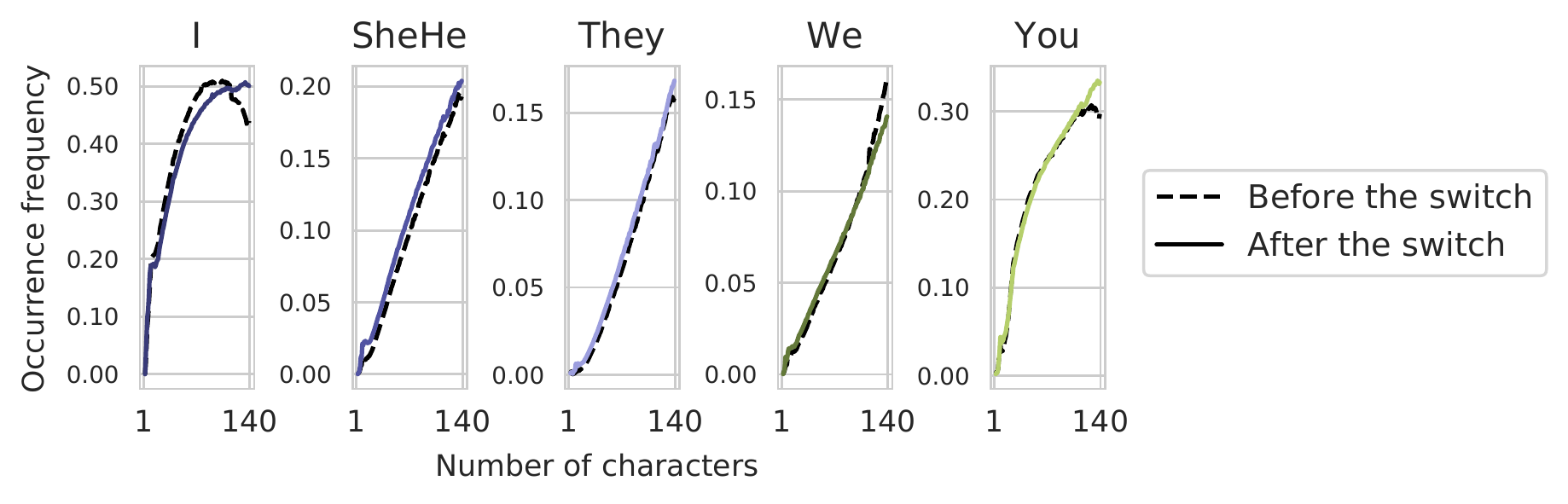}
    \subcaption{Before \vs\ after the switch}
    \label{fig:12}
    \end{minipage}
    \hfill
    \begin{minipage}[t]{.95\textwidth}
    
    \hspace{15pt}
    \includegraphics[width = 0.66\textwidth]{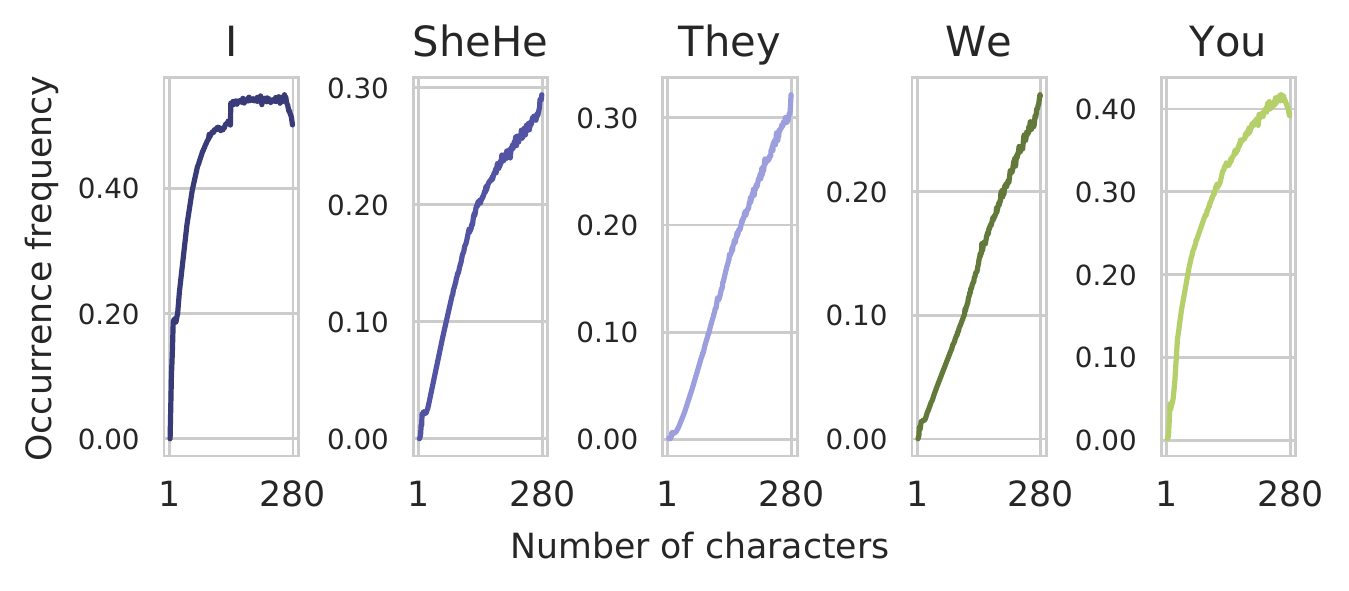}
    \subcaption{After the switch}
    \label{fig:13}
    \end{minipage}
\caption{\textbf{Syntactic indicators of cramming: personal pronoun frequency across tweet lengths.} (a)~Occurrence frequency of personal pronouns across tweets of different character lengths in the period before the switch (1--140 characters). The dashed black line represents this quantity across tweets posted in the period before the switch (\ie, under the 140\hyp character limit), and the solid colored line in the period after the switch (\ie, under the 280\hyp character limit).
(b)~Occurrence frequency of personal pronouns across tweets posted in the period after the switch (\ie, under the 280\hyp character limit) for different lengths (1--280 characters). Personal pronouns are sorted alphabetically. Note the different $y$-axis scales.
}
\label{pronouns}
\end{figure}

\subsubsection{Gaps across languages} 

These insights are robust across languages (\Figref{panel1} and \Figref{panel2}). The actual usage of long tweets surpassed estimated cramming the most in Hindi, Urdu, and Russian (\Figref{panel1}). Cramming at 280 characters, smaller in size compared to cramming at 140 characters, and more pronounced on the Web interface compared to the mobile devices, emerged in all languages (\Figref{panel2}).

\subsubsection{Syntactic indicators of cramming} 
In what follows, we aim to understand further the cramming that emerged at 280 characters after the switch by studying its impact on linguistic features of the tweets. Is cramming also evident in the text of tweets?
Why do users post long tweets? What are their signature characteristics, and are they indicative of cramming as revealed via length modeling?

To answer these questions, we study tweets in English posted from mobile and Web devices during the pre- (75.56M tweets) and post-switch (65.29M tweets) periods. We annotated the tweets with LIWC's \cite{pennebaker2007liwc2007} syntactic features (linguistic categories) and semantic features (psychological, biological, and social categories).

First, to characterize syntactic features of tweets, for all tweets with a given number of characters, we measure the occurrence frequency of part of speech (POS) tags among tweets of that length. In \Figref{fig:10}, across all possible tweet lengths in the period before the switch (1--140 characters), we observe the fraction of tweets of that length that have at least one instance of the respective POS tag. The dashed black line represents this quantity across tweets posted in the period before the switch (\ie, under the 140\hyp character limit), and the solid colored line represents this quantity among tweets posted in the period after the switch (\ie, under the 280\hyp character limit). 

Comparing the solid colored and dashed black lines across different POS tags lets us isolate the effect of the length limit on the content of tweets. The largest gap is observed for swear words and spoken categories (nonfluencies, fillers, and assent), adverbs, and conjunctions---nonessential parts of speech that are frequently deleted in the process of ``squeezing'' a message to fit a length limit. No gap is observed for verbs and negations, essential parts of speech that are known to be disproportionately preserved in the cramming process~\cite{10.1145/3359147}.

\begin{figure}[t]
    \begin{minipage}{.49\textwidth}
        \centering
        \includegraphics[width=\textwidth]{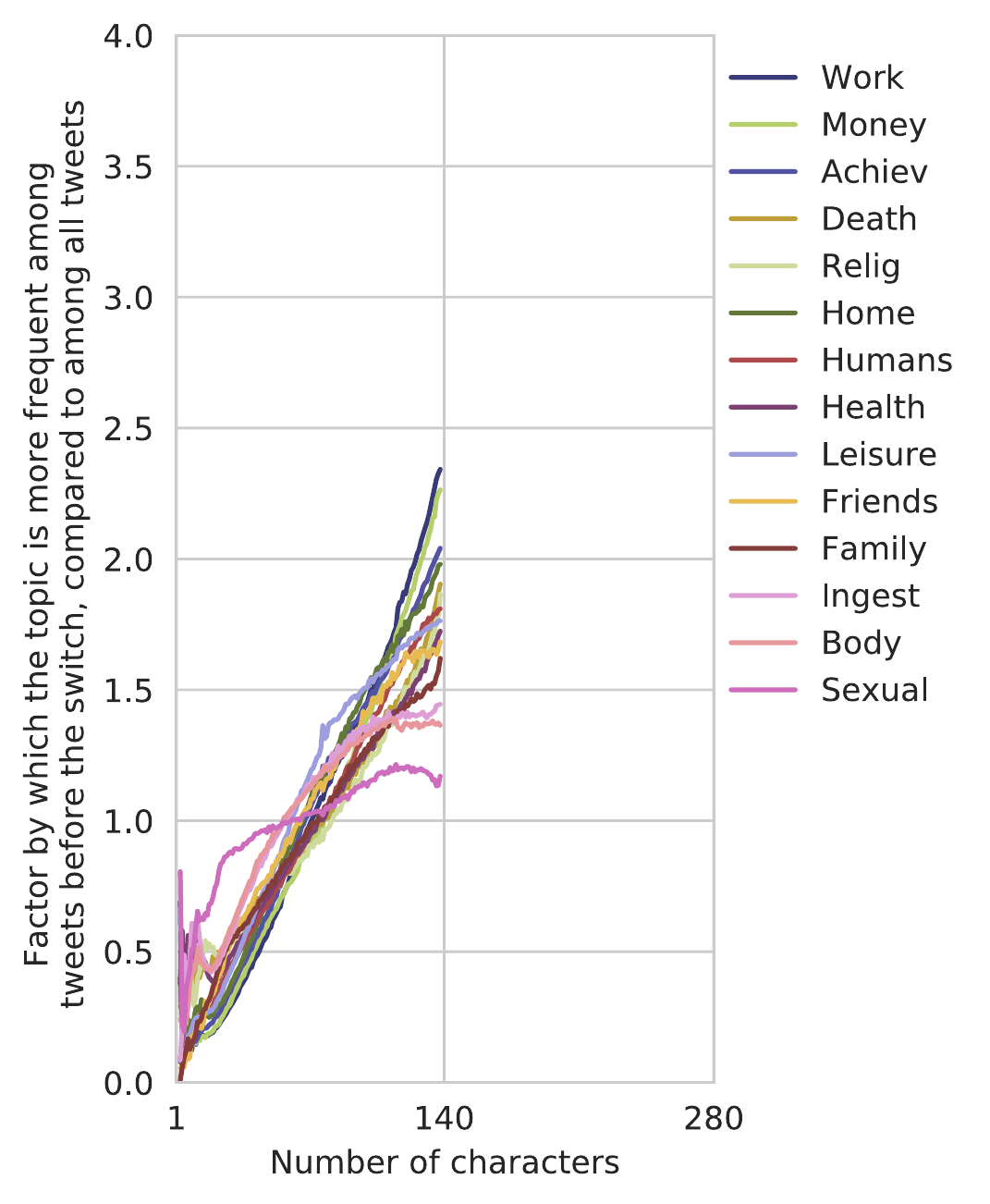}
        \label{fig:14}
    \end{minipage}
    \begin{minipage}{.49\textwidth}
        \centering
        \includegraphics[width=\textwidth]{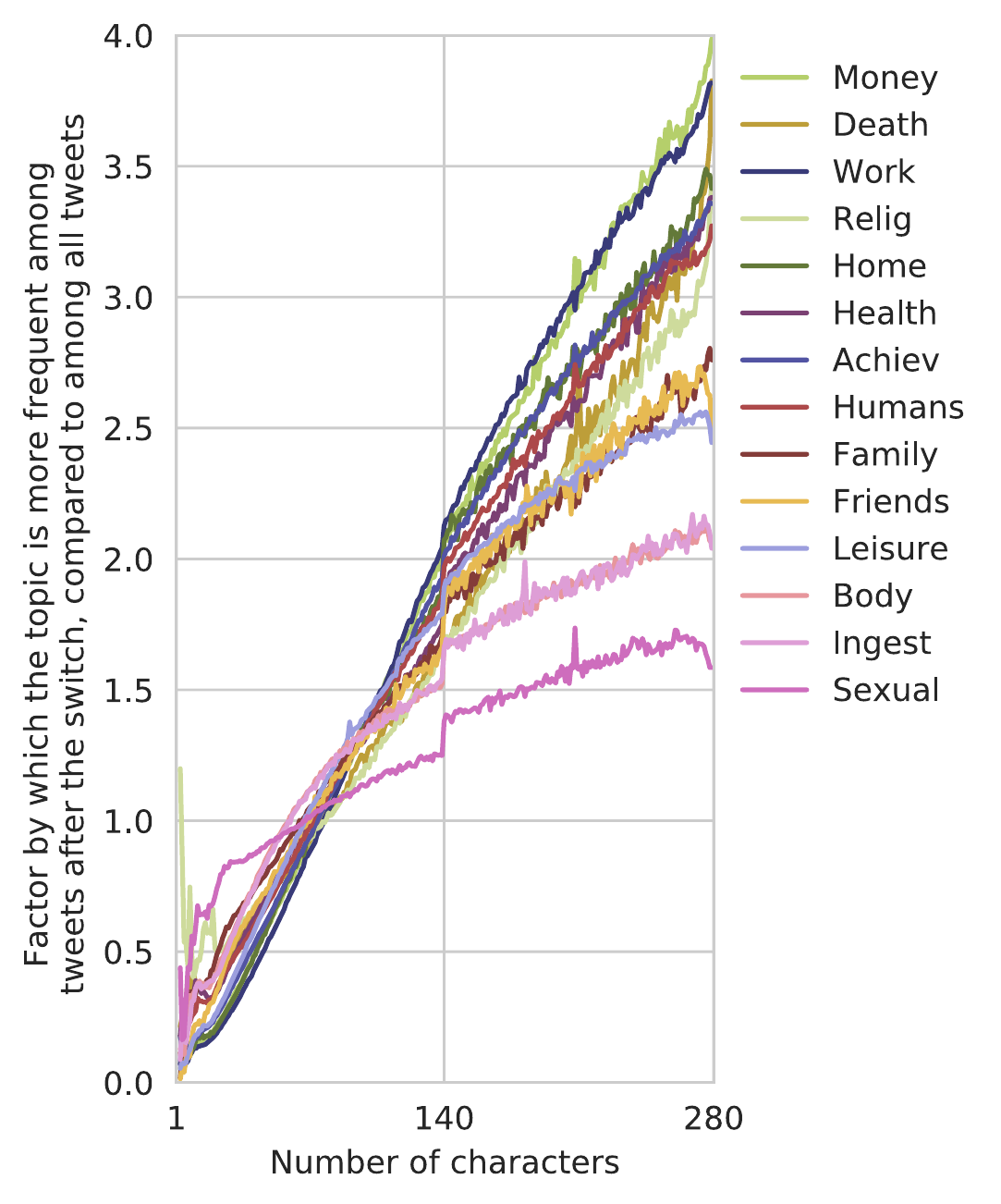}
        \label{fig:15}
    \end{minipage}
\caption{\textbf{Semantic indicators of cramming: topics across tweet lengths.} For each length before the switch (left) and after the switch (right), across 14 topics measured with LIWC categories \cite{pennebaker2007liwc2007}, we monitor the factor by which the topic is more frequent among tweets with that number of characters, compared to tweets across all lengths. Categories are sorted by the value of this factor at 140 characters (left) or 280 characters (right).}
\label{fig:politics}
\end{figure}

\Figref{fig:11} represents the same quantities after the switch (\ie, under the 280\hyp character limit) across all possible tweet lengths in this period (1--280 characters). We note that there is no counterfactual observation here, \ie, we do not know what the probability of observing a POS tag among 280\hyp character tweets would be if the 280\hyp character limit was lifted. Nonetheless, among the 280\hyp character tweets after the switch, we do observe patterns similar to those associated with the 140\hyp character tweets before the switch, in particular, a dip in the frequency of the spoken categories, conjunctions, and numbers among the 280\hyp character tweets, traces typical of cramming or ``optimizing'' a message to fit a length limit.

In \Figref{fig:12} and~\Figref{fig:13} we measure the same quantities for fine-grained subtypes of personal pronouns. This suggests that personal pronouns \textit{I} and \textit{you} were most affected by the 140\hyp character limit (\ie, they were most likely to be omitted before the switch), and we observe a similar non-monotonic pattern around 280 characters in \Figref{fig:13}.

To summarize, 280\hyp character tweets from after the switch are \emph{syntactically} similar to 140\hyp character tweets from before the switch, following patterns indicative of cramming~\cite{10.1145/3359147}. This is evidence indicating that they are generated by similar writing processes as 140\hyp character tweets used to be, and provides further evidence of emerging cramming behavior at the new length limit.

\subsubsection{Semantic indicators of cramming}

In \Figref{fig:politics} we next explore topics of tweets, as measured by LIWC categories describing psychological, biological, and social categories. Across the studied topics, for each allowed tweet length in the pre-switch and post-switch periods, we measure the factor by which the topic is more frequent among tweets of a given length, compared to all tweets across lengths. Categories are sorted by the value of this factor at 140 characters before the switch, and at 280 characters after the switch.

We note that the personal-concerns categories were relatively most frequent around 140 characters before the switch: \textit{Work}, \textit{Money}, \textit{Achievement}, \textit{Death}, and \textit{Religion}. The least frequent topics at 140 characters, on the other hand, were the biological categories: \textit{Sexual}, \textit{Body}, \textit{Ingestion}, followed by \textit{Family}, \textit{Friends}, and \textit{Leisure}. An apparent association with the importance of the message emerges: while tweets about topics related to ordinary, overall more prevalent everyday experiences are long the least frequently, topics related to more serious personal concerns are long most frequently. There is a within-topic correlation between usage of 140 characters before the switch, and subsequent usage of 280 character length after the switch, with the ranking of topic usage at the boundary length only slightly changed (Spearman's rank correlation between topics $0.91$, $p = 7.30 \times 10^{-6}$), implying that 280\hyp character tweets from after the switch are also semantically similar to 140\hyp tweets from before the switch.
This is further evidence indicating that they are generated by similar processes as 140\hyp character tweets used to be.

 \begin{figure}
     \centering
     \includegraphics[width= 0.99\textwidth]{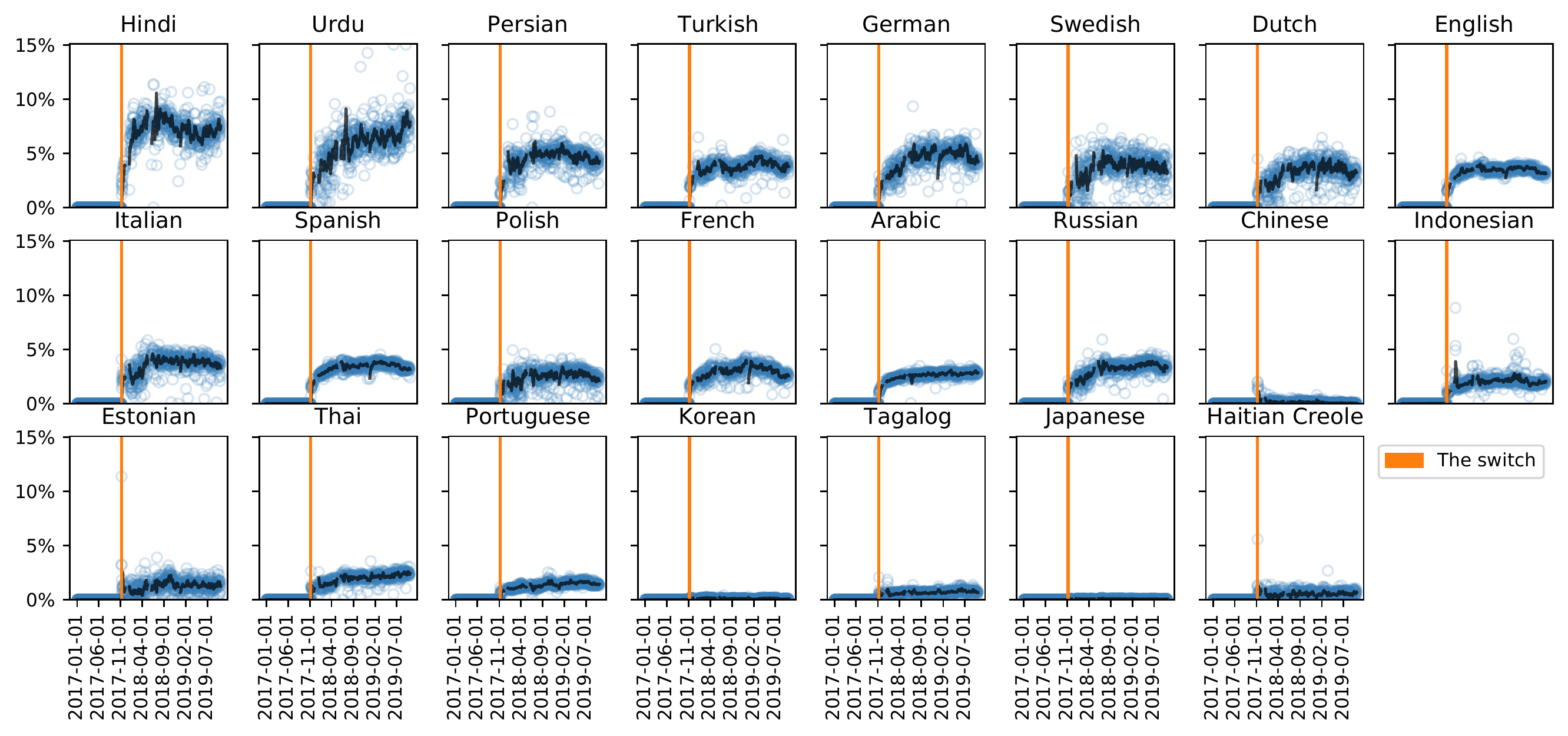}
     \caption{\textbf{Daily evolution of the estimated size of cramming at 280 characters, across languages.} Daily fraction (indicated with a circle) and 10-day rolling average (solid line) of the estimated size of cramming. Languages sorted by size of cramming at 140 characters before the switch.}
     \label{figlangs1}
 \end{figure}

\subsection{RQ2: Cramming across languages}

Given that the switch constitutes a global platform design change that affected widely different user populations, we aim to further understand its impact across the globe.
In particular,
we study the temporal evolution of cramming and contrast how different user populations are affected across languages.

In \Figref{figlangs1}, we monitor the evolution of cramming at the new limit across languages, for tweets posted from Web and mobile devices. In most of the languages, the cramming seems to have settled, and is even decreasing again in some languages. Urdu is a notable exception, where the size of cramming is still growing and not yet in a stable state.

\begin{figure}
    \centering
    \includegraphics[width= 0.9\textwidth]{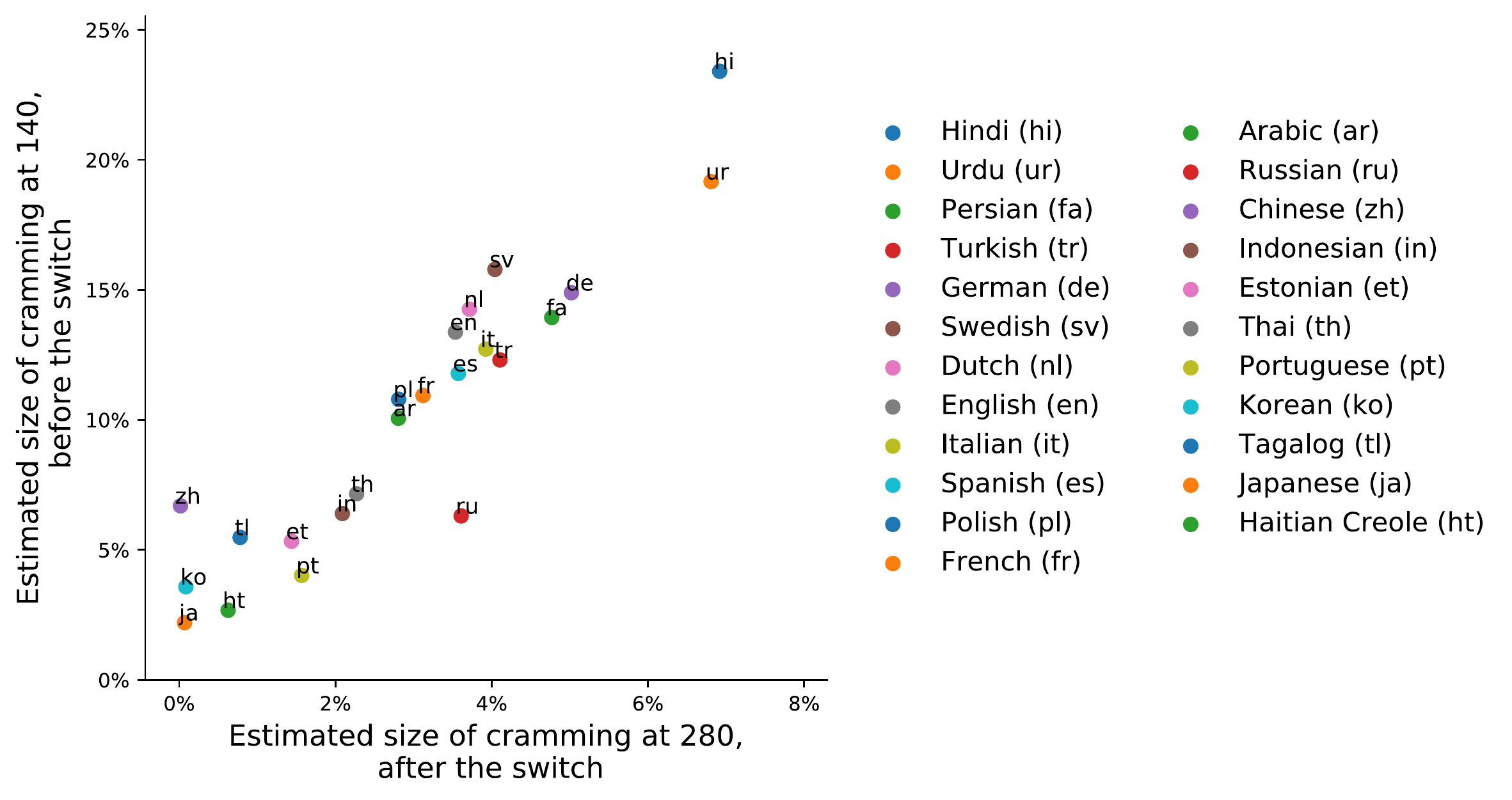}
    \caption{\textbf{Cramming in 23 languages before \vs\ after the switch.} Estimated size of cramming at 280 characters after the switch ($x$-axis) \vs\ estimated size of cramming at 140 characters before the switch ($y$-axis) (Spearman's rank correlation $0.90$, $p = 5.47 \times 10^{-9}$).}
    \label{figlangs2}
\end{figure}

In \Figref{figlangs2}, we show the estimated size of cramming at 280 characters after the switch ($x$-axis) \vs\ the estimated size of cramming at 140 characters before the switch ($y$-axis). We observe that the estimated size of cramming at 140 before the switch in a language is highly correlated with the estimated size of cramming at 280 after the switch: the more cramming there was in a language at 140 before the switch, the more cramming there is at 280 after the switch (Spearman's rank correlation $0.90$, $p = 5.47 \times 10^{-9}$). In Hindi and Urdu, there is particularly much cramming at 280 after the intervention, to such an extent that 280 characters is the most frequent tweet length after the switch (\Figref{fig:histlangs}).

In summary, we observe that the more cramming there was at 140 in a language before the switch, the more cramming there is at 280 after (\Figref{figlangs2}). Disaggregation across languages reveals interesting temporal patterns (\Figref{figlangs1}), whereby in most languages the cramming seems to have settled and is even decreasing again, \ie, the peak of cramming is over.

\subsection{RQ3: Fluidity of counterfactual estimates}

\subsubsection{Tweets that fit the new limit}

The new policy (280 characters) was selected because it was estimated that if that limit were enforced, a negligibly low fraction of tweets would not fit the limit~\cite{t3}. However, as examined up to this point, cramming emerged again at the new limit. Hence, we explore how the estimated impact of the character limit evolves as user behavior evolves, using the same modeling approach applied by Twitter before the switch.

\begin{figure}[t]
    \centering
    \includegraphics[width= 0.9\textwidth]{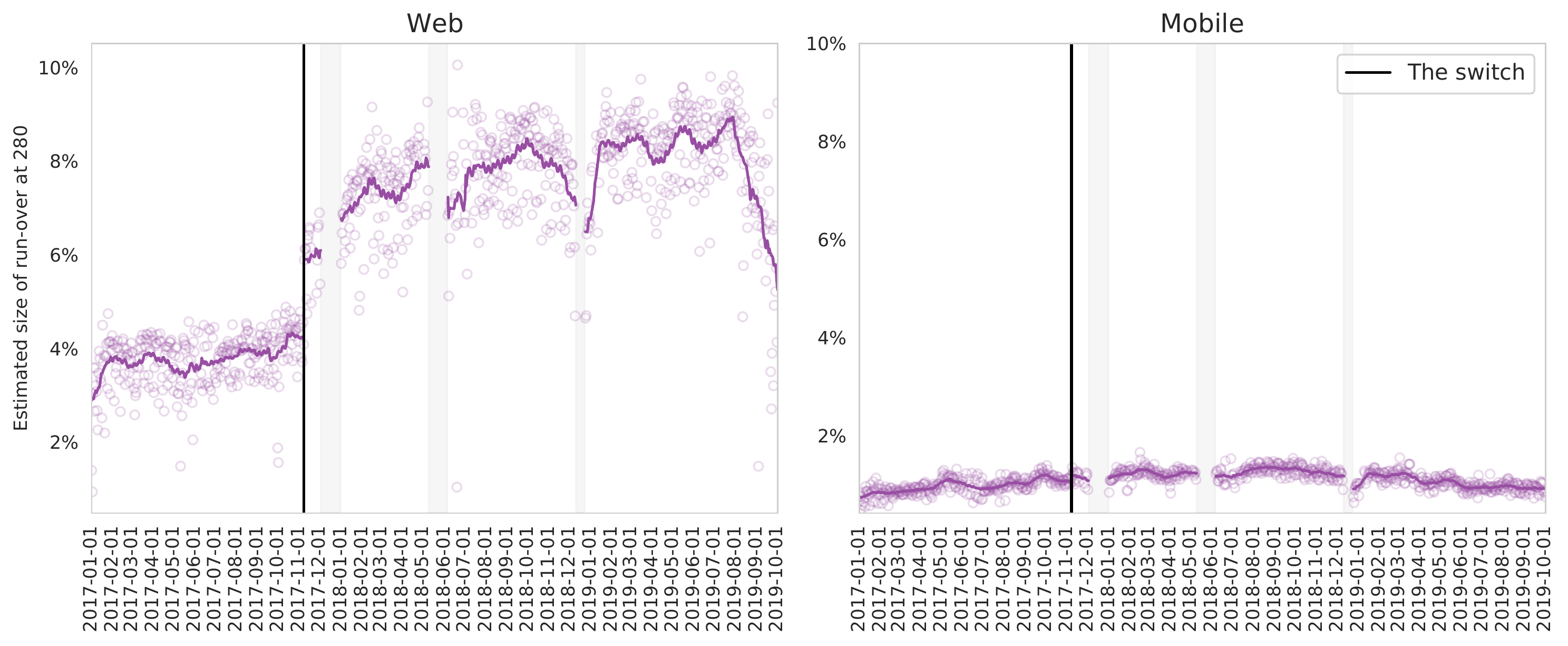}
    \caption{\textbf{Evolution of estimated size of run\hyp over at 280 characters.} Circles indicate the daily estimated fraction of tweets that would be longer than 280 characters if it were possible; the line marks the 10-day rolling average. The quantities are shown separately for the Web interface (left) and mobile applications (right). The vertical line marks the switch, and gray bands mark days with missing data. After the switch, the estimated size of run\hyp over at 280 characters increases sharply on the Web interface, while it increases only slightly on mobile devices.}
    \label{runover1}
\end{figure}

In \Figref{runover1}, we measure the estimated size of run\hyp over at 280 characters. In the case of the Web interface before the switch, 3.76\% [3.69\%, 3.83\%] of tweets were estimated to be longer than 280 characters if it were possible, while after the switch, this number increases to 7.77\% [7.68\%, 7.86\%].
In the case of mobile devices before the switch, it was estimated that 0.97\% [0.96\%, 0.99\%] of tweets would be longer than 280 characters if it were possible. After the switch, this number remains close, at 1.15\% [1.14\%, 1.16\%].

In summary, the estimated size of run-over at 280 characters more than doubled on Web interface (+107\%) and slightly increased on mobile devices (+19\%).
Now, more tweets would be longer than 280 if they could be, implying that there is fluidity of estimates due to the fluidity of user behavior---both impacted by the currently imposed limit.

\subsubsection{Necessary number of characters to fit all tweets under the new limit}

Finally, we apply the same reasoning used before the switch and ask: For a targeted size of run-over, what character limit should be chosen? How many characters are necessary for a targeted fraction of tweets to fit the new limit? What future policies would reduce the cramming that emerged at the new character limit? If a new character limit were to be imposed to reduce cramming at 280, how many characters would it need to be?

\begin{figure}[t]
    \centering
    \includegraphics[width= 0.9\textwidth]{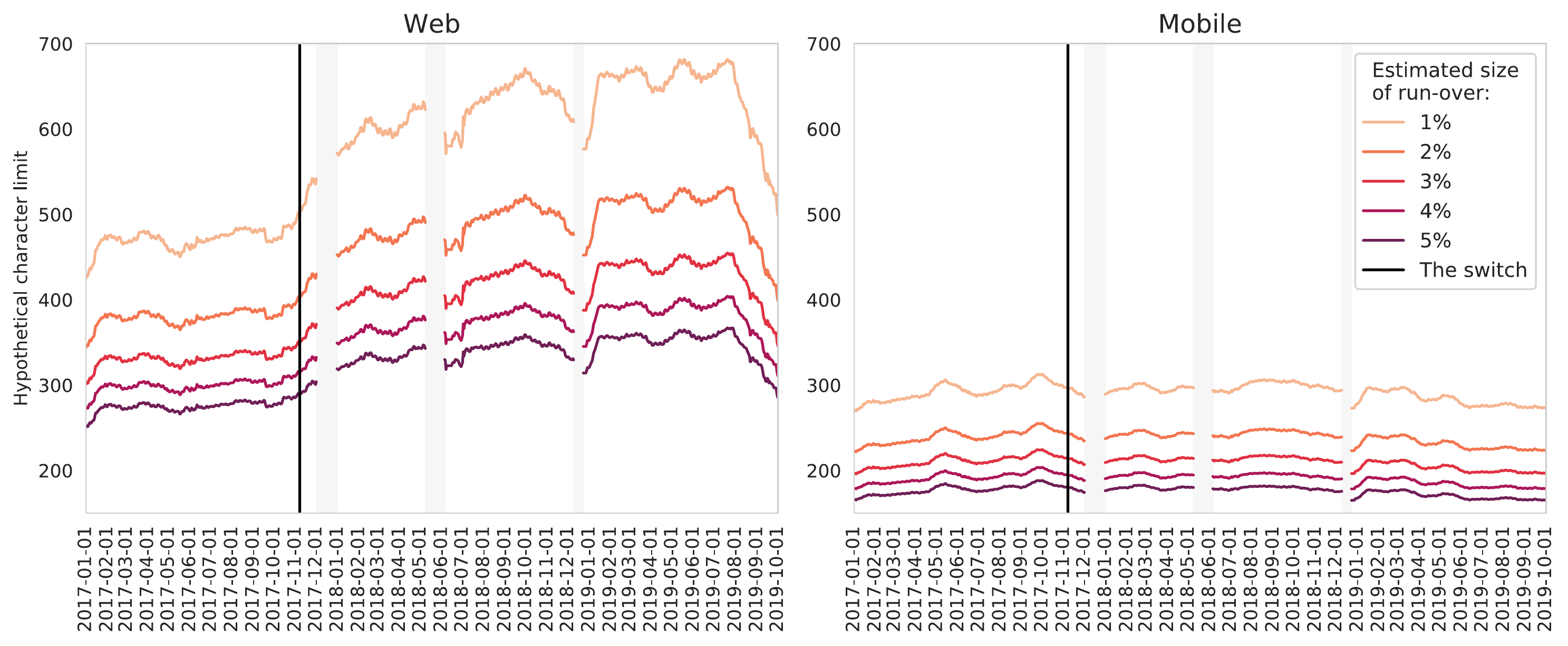}
    \caption{
    \textbf{Evolution of hypothetical character limit necessary to achieve various targeted sizes of run\hyp over.} Tweet length limits (in the number of characters) are shown on the y\hyp axis; targeted sizes of run\hyp over are marked in colors. Lines mark 10-day rolling averages. The quantities are shown separately for Web interface (left) and mobile applications (right). The vertical line marks the switch and gray bands mark days with missing data. After the switch, the number of characters needed to achieve a targeted run\hyp over increases sharply on the Web interface, while it remains robust on the mobile devices.}
    \label{runover2}
\end{figure}

In \Figref{runover2}, we show the hypothetical tweet length limits necessary to achieve various targeted sizes of run\hyp over. For instance, to make 95\% of tweets fit (that is, to achieve 5\% of run-over) on the Web interface before the switch, an estimated 275 (95\% CI [273, 277]) characters would be necessary. After the switch, the number of characters would need to be increased to 342 [340, 344]. On mobile devices, there is no significant increase in the necessary number of characters to fit  95\% of tweets (178 [177, 179] characters before \vs\ 175 [174, 177] characters after).
Note that these estimates do not take into account the dynamic, fluid nature of user behavior in response to design change, but take the simpler, static view instead, mimicking the view taken by Twitter in their decision\hyp making before the switch.

If, after the switch, 1\% run-over (or fitting 99\% of tweets) on the Web interface were desired, as many as 628 [623, 633] characters would be necessary. It is likely that, no matter how much one would increase the number of characters, the new limit would again lead to cramming, and the number of characters would have to be further increased to make more tweets fit the new limit.

Cramming is inherent to writing text under a character limit. When anticipating the impact of interventions, it is not sufficient to estimate how many tweets would be longer if the current limit did not exist (\eg, by modeling the run\hyp over). It is necessary to also account for new cramming that would emerge given any other hypothetical limit.

%% file: 5disscusion.tex
\subsection{Implications for the design of socio\hyp technical systems and future platform changes}

In this work, we analyze the aftermath of a design decision that shaped a socio\hyp technical system and impacted millions of users. Twitter engineers and policy\hyp makers modeled historical user behavior traces in order to design the platform intervention. By doing so, it was implicitly assumed that the user behavior would not in other ways change in response to the implementation of the intervention. We find that the actual user behavior diverged from anticipated behavior since cramming emerged at the new character limit, and the usage of long tweets turned out to be higher than predicted (\Figref{timeseries}). Such a user response was apparently not taken into account when anticipating the effect of the intervention. Our findings suggest the need to account for user response and highlight the fluidity of online behaviors. 

The new length limit (280 characters) was selected since it was estimated that, if that limit were enforced, a negligibly low fraction of tweets would be impacted by cramming~\cite{t3}. After the intervention, using the same modeling methodology, we estimate that a further increase of the allowed number of characters would be necessary to achieve the same goal (\Figref{runover2}). The evolution of this counterfactual estimate speaks about the fluidity of predictions and the limits to the predictability of user behaviors. The estimated effect of a policy changes as user response is taken into account. \minor{These findings fill the gaps between the literature on strategic feedback that takes into account the feedback of the environment \cite{hardt2016strategic,miller2020strategic} and the static user modeling practices outlined in \Secref{sec:relfeedback}. Our findings reveal that although a static view is often assumed in practice, in the case of Twitter's character limit change, it was not justified. The analyses of the user response highlight that, moving forward, it is necessary to account for such responses of the users and the environment by developing novel dynamic and integrative user modeling approaches.}

The major challenge of anticipating the impact of platform design changes is considering and incorporating user reactions.
To model design change in the presence of users' adapting their behavior in response, more advanced, game-theoretic approaches may be required.
Note that even experimental approaches such as A\slash B tests might fall short if not performed longitudinally over longer periods of time, allowing for users to adapt their behavior in response.
An additional promising way forward may be to integrate sensitivity bounds into the estimates, in the spirit of how sensitivity analysis of a causal estimate is performed \cite{pearl2018book,rosenbaum2005sensitivity,rosenbaum2010design}. It is helpful to consider in advance how severe the users' dynamic response would need to be such that the predictions do not carry any statistical significance anymore. Domain knowledge can then be incorporated to reason about the plausibility of such a sufficiently drastic user response, ahead of the intervention.

\subsection{Implications for Twitter research}

Although the doubling of the length limit reduced the amount of cramming and eliminated the drastic disproportion of tweets reaching the maximum length (\eg, 9\% of English tweets used to be exactly 140 characters long before the switch), our results demonstrate the emergence of a similar, though considerably weaker, effect around 280 characters after the switch. Hence the new character limit can be seen as a less intrusive version of the previous 140 character limit.

These findings have important implications for Twitter\hyp based research, as they show that, although the new limit is ``felt less'' by users than the old limit, 280 characters still constitutes an impactful length constraint that shapes the nature of Twitter. The evidence suggests that, just as the old 140\hyp character limit \cite{w1}, the new 280\hyp character limit impacts the writing style (\Figref{pos_tags} and \Figref{pronouns}) and content of tweets (\Figref{fig:politics}) via cramming \cite{sen2019total}. The length constraint and the resulting tweet\hyp length distribution remain an important dimension to consider in studies using Twitter data \cite{olteanu2019social}. After the switch, the number of characters remains an important variable, correlated with important user features including device, language, and topics.
In a nutshell, ``280 is the new 140'', although it is less intrusive.

\subsection{Differences across devices}

We observe widely different patterns between Web and mobile devices (\Figref{timeseries}). On the Web interface, user behavior widely diverged from what had been anticipated, and considerable cramming appeared at the new limit. On mobile devices, user behavior remained close to the predictions made before the intervention, and little cramming emerged at the new limit. We highlight this distinction between Web and mobile devices. This bimodal nature of Twitter as a platform should be carefully taken into account in future studies of online platforms.

\rev{Cramming and usage of long tweets are particularly prominent on Web clients. The fact that tweets were longer on the Web interface before the switch also indicates a tendency for shorter text on mobile phones. The degree to which the long tweet length is consistent with experiences and needs of users writing tweets can potentially explain the differences between Web and mobile devices. Typing shorter texts remains more compatible with small touchscreens on mobile devices \cite{buschek2015improving}. Hence, users on mobile devices may experience less the need to ``squeeze'' their message.}


\subsection{Syntactic and semantic characteristics}

Tweets of 280 characters are syntactically similar to 140\hyp character tweets before the switch, following patterns indicative of cramming a message (Figures~~\ref{fig:10} and~\ref{fig:11}). This is evidence indicating that tweets close to the new boundary are generated by similar writing processes as 140\hyp character tweets were before the switch. There is a within-topic correlation between usage of 140 characters before the switch, and subsequent usage of 280 characters after the switch (\Figref{fig:politics}). Beyond syntax, tweets of 280 characters are also semantically similar to 140\hyp character tweets before the switch.
While tweets about topics related to ordinary, overall more prevalent everyday experiences use the longer tweets the least frequently, topics related to more serious personal concerns use them the most. This apparent association with the importance of the message is aligned with previous work that described how Twitter's signature feature---brevity---is particularly pronounced in the context of political expression \cite{priya2019should} and mental\hyp health discussions \cite{de2014mental}, since longer messages are used for more complex sentences and ideas \cite{halpern2013social}. The fact that ``important'' topics were affected by cramming more than other topics before the switch, and are affected more after the switch, highlights the need to understand the implications of the implemented character limit intervention. Further considerations of policy changes on social platforms should take into account the effect they have on socially important discourses such as politics \cite{huszar2022algorithmic}. \minor{These findings fill gaps in the literature on Twitter communication and supporting features outlined in \Secref{sec:relcomm}. In particular, while most of the previous work studied how users adapt to new features such as retweets, hashtags, and quotes in the short term \cite{boyd2010tweet,wikstrom2014srynotfunny,page2012linguistics,garimella2016quote,pavalanathan2016more}, our results highlight the importance of analyzing the response in the long run. Our findings also fill gaps in the Twitter communication literature \cite{shapp2014variation,ciot2013gender,sylwester2015twitter} by providing a novel characterization of adoption heterogeneity across topics and user subpopulations.}

\subsection{Differences across languages}

The language-specific findings reveal notable differences between languages that were not a priori expected. Hindi and Urdu are the languages where there was most cramming at 140 characters before the switch, and where there is the most cramming at 280 characters after the switch (\Figref{figlangs1}). These two languages, additionally, exhibit a different cramming evolution pattern compared to other languages, with the size of cramming still not in a stable state, but growing. This raises the question: What makes the cramming size different in those two languages? It is interesting to note that Hindi and Urdu are mutually intelligible as spoken languages. However, they are written in different scripts: Devanagari and a Perso-Arabic script, completely illegible to readers literate only in one of the two.

\rev{A possible explanation could be the fact that some languages might need more or fewer characters to express the same amount of information \cite{coupe2019different}. The fact that we observe that the cramming at 140 characters before the switch in a language is correlated with cramming at 280 after the switch potentially points to information density remaining constant in a given language. Beyond possible language-specific reasons, cultural and broader societal values in complex ways influence the use of technology \cite{wejnert2002integrating}. Future work should understand better the factor driving the observed differences between languages.}


\subsection{Limitations} 

\rev{This study suffers from limitations that the 1\% stream of tweets is known to be susceptible to \cite{wu2020variation}, as certain accounts might be over-represented due to the intentional or unintentional tampering with the Sample API
\cite{pfeffer2018tampering,morstatter2013sample}. Due to the nature of Twitter’s sampling mechanism, it is possible to deliberately influence the studied sample, the extent and content of any topic. Additionally, technical artifacts can skew Twitter’s samples. Therefore, the 1\% stream of tweets cannot be regarded as fully random \cite{pfeffer2018tampering}.} Additionally, in our study, we focus on the most common sources of tweets: the Web interface and mobile applications. We disregard automated sources and third-party applications as a proxy for bots. However, bot detection can be more reliable using more sophisticated methods detecting bots that use regular applications \cite{chavoshi2016debot,kudugunta2018deep,yang2020scalable}, which we did not consider in this study.

We note that our syntactic and semantic analysis of tweets is limited to tweets in English only, due to the lack of available tools to support annotation consistently across all the studied languages. Future studies should measure these characteristics in other languages, using language-specific tools or machine translation. Finally, we note that in this study, we study Twitter as a platform (tweets are sampled at the community level), as opposed to users, whose timelines are incomplete in the 1\% sample. 

\subsection{Future work}

Future work should provide a better understanding of what user-specific features are associated with cramming behavior. Twitter hoped that increasing the length limit would reduce the friction to tweet and that changing the limit might therefore increase the fraction of users who post \cite{t3}. Future work should determine whether more users indeed tweeted due to the character limit intervention. Furthermore, future work should develop novel approaches for better modeling how design changes will impact online platforms. For instance, beyond static modeling based on historical data, future work might consider incorporating a game\hyp theoretic analysis in order to anticipate how users might respond to platform design changes.

In the future, one could also focus on the question of age and ask: Are the users who have long been cramming under 140 the same ones who are more likely to cram under 280? Answering these questions requires data beyond the 1\% sample, with complete records of users' tweets. However, here we caution against na\"ive comparisons, as careful quasi-experimental designs are necessary to truly isolate the effect of platform change and the effect of age on specific users \cite{barbosa2016averaging}. User activity\hyp level and age are correlated with other factors; \eg, users who stay longer on the platform might be in other ways fundamentally different from younger users who joined more recently (\ie, there is ``survivor bias'' \cite{elton1996survivor}). Finally, our study should be replicated in the future as new platform design changes are implemented.

\subsection{Ethical considerations}

We perform a population-level study of user behavior that does not focus on any individual user. All analyses are based on highly aggregated daily statistics.

\subsection{Code and data}

Code and data necessary to reproduce our results at are publicly available at: \url{https://github.com/epfl-dlab/anticipated-vs-actual}.

%% file: 6conclusion.tex
Reflecting on our main research question, we find evidence that after the introduction of the 280\hyp character limit, a gap between anticipated and actual behavior emerged. After the intervention, users started using long tweets more than anticipated and cramming emerged at the new length limit. Cramming is likely inherent to producing text under any length constraint and, as such, should always be taken into account when designing platforms that allow the production of textual content. The user modeling approach used by Twitter did not consider such shifts in user behavior as a response to the platform change. \rev{Moving forward, when anticipating the effect of platform changes, cautious approaches that aim to consider the dynamic interaction between platform design and user behavior, as well as the impact the change will have on different populations depending on users' languages and devices, are necessary.}

%% file: 7appendix.tex
\subsection{Threaded tweets}

Our main analyses study tweets in isolation. However, Twitter users have long been working around the character limit by splitting a long piece of text into a sequence of length-compliant tweets. These tweet threads are usually connected with the supporting threading feature, or annotated with the position of the tweet in its sequence. They are not intentionally altered to fit the character limit requirement. We investigated user behavior a step beyond the single tweets to estimate the impact of the introduction of 280 character length on the length of threads connecting tweets.

For each thread length $k$, we estimate the expected number $n_k$ of tweets from threads of length $k$ as the empirical number of tweets ending with the conventional pagination ``$i$/$k$'':
\begin{equation}
    n_k = \epsilon \cdot k \cdot m_k,
    \label{eqn:formula_threads1}
\end{equation}
where $\epsilon= 0.01$ is the sampling rate, and $m_k$ is the total number of threads of length $k$. With this, we can estimate the total number of threads of length $k$ as
\begin{equation}
    m_k = \frac{n_k}{\epsilon \cdot k}.
    \label{eqn:formula_threads2}
\end{equation}

\begin{figure}[b]
    \centering
    \includegraphics[width= 0.45\textwidth]{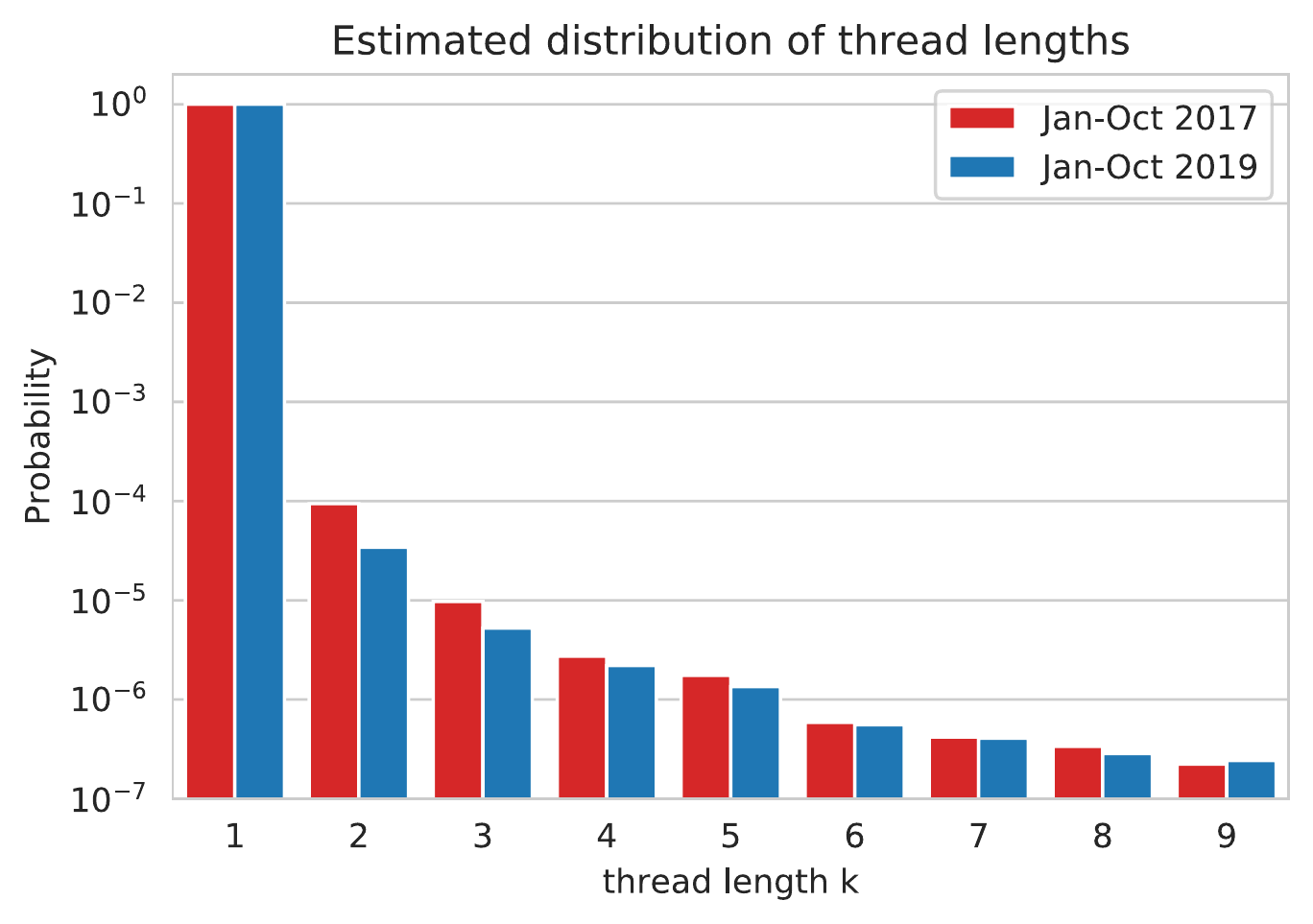}
    \caption{\textbf{Threaded tweets, before \vs\ after the switch.} Estimated distribution of thread lengths, before the 280 limit was introduced (in red), and after it was introduced (in blue).}
    \label{fig_threads}
\end{figure}

We then estimate the distribution of thread lengths before and after the 280 character limit was introduced (Figure~\ref{fig_threads}), for tweets tweeted in one of the 20 studied languages where the switch occurred and posted from Web and mobile agents. Tweets without pagination are considered as threads of length $k=1$. In Figure~\ref{fig_threads}, we depict the estimated distribution of thread lengths, before the 280 limit was introduced, and after.

Although threading is not a prevalent behavior---fewer than 0.1\% of all tweets are paginated \cite{w1}---, there is evidence of less splitting of long texts into sequences of length-compliant tweets after the introduction of the new limit. We observe that threaded tweets occurred more frequently before the 280 limit was introduced and that long threads were more common before the introduction. However, there are no drastic shifts in the distribution of thread lengths due to the introduction of the 280 limit. Hence, threading is not accounted for when modeling cramming and run\hyp over in our main analyses.

\subsection{The contrasting case of Chinese, Japanese, and Korean}

Additionally, we take advantage of the fact that the new character limit was not introduced in all languages to perform a differences in differences estimation of the effect of the switch on tweet lengths.

To account for possible global platform-wide changes that are not associated with the switch, we use a differences in differences regression estimation, where the tweet lengths in Japanese, Korean and Chinese, languages where the 280 characters were not introduced are the control time series, and the tweet lengths in the other 20 studied languages (Figure ~\ref{fig:counts}) are the treated time series. Both are observed in the pre-switch (Jan-Oct 2017), and post-switch (Jan-Oct 2019) periods.
We fit a model
\begin{equation}
    y \sim \mathtt{treated * period},
    \label{eqn:formula_overall}
\end{equation}
where the dependent variable $y$ is the logarithm of the average tweet length for each studied calendar day,
and as independent variables are the following two factors:
$\mathtt{treated}$ (indicates whether the switch was introduced or not in those languages),
$\mathtt{period}$ (indicates whether a calendar day is in year pre-switch or post-switch). 
$\mathtt{treated * period}$ is shorthand notation for $\alpha + \beta \mathtt{treated} + \gamma \mathtt{period} + \delta \mathtt{treated:period} + \epsilon$, where in turn $\mathtt{treated:period}$ stands for the interaction of $\mathtt{treated}$ and $\mathtt{period}$.

The interaction term $\mathtt{treated:period}$ $\delta$ is then the effect of switch on the logarithm of average tweet length. Each studied pre- or post-switch period spans 277 days per condition, amounting to a total of $4 \times 277 =$ 1108 data points. The model is multiplicative due to the log. The relative increase over the baseline is then calculated by converting back to the linear scale the fitted coefficient $\delta$. Fitting the model~\ref{eqn:formula_overall}, we measure a $e^{\delta}-1 = e^{0.0598}-1 = 6.16\%$ (95\% CI [5.68\%, 6.64\%]) increase in tweet lengths in the languages where the switch happened, over the control baseline. We note, however, that tweet lengths increased slightly in the control languages as well. This is likely impacted by the nature of how tweet length is counted at the character level, allowing mixed\hyp character tweets to be longer than 140 characters. 

To summarize, we estimate a significant increase in tweet lengths in languages where the new character limit was introduced, compared to the control languages, thus accounting for possible global platform-wide changes that are not associated with the doubling of the character limit intervention.

\subsection{Gaps between predictions and emerging user behavior across languages}

\begin{figure}[b]
    \centering
    \includegraphics[width= \textwidth]{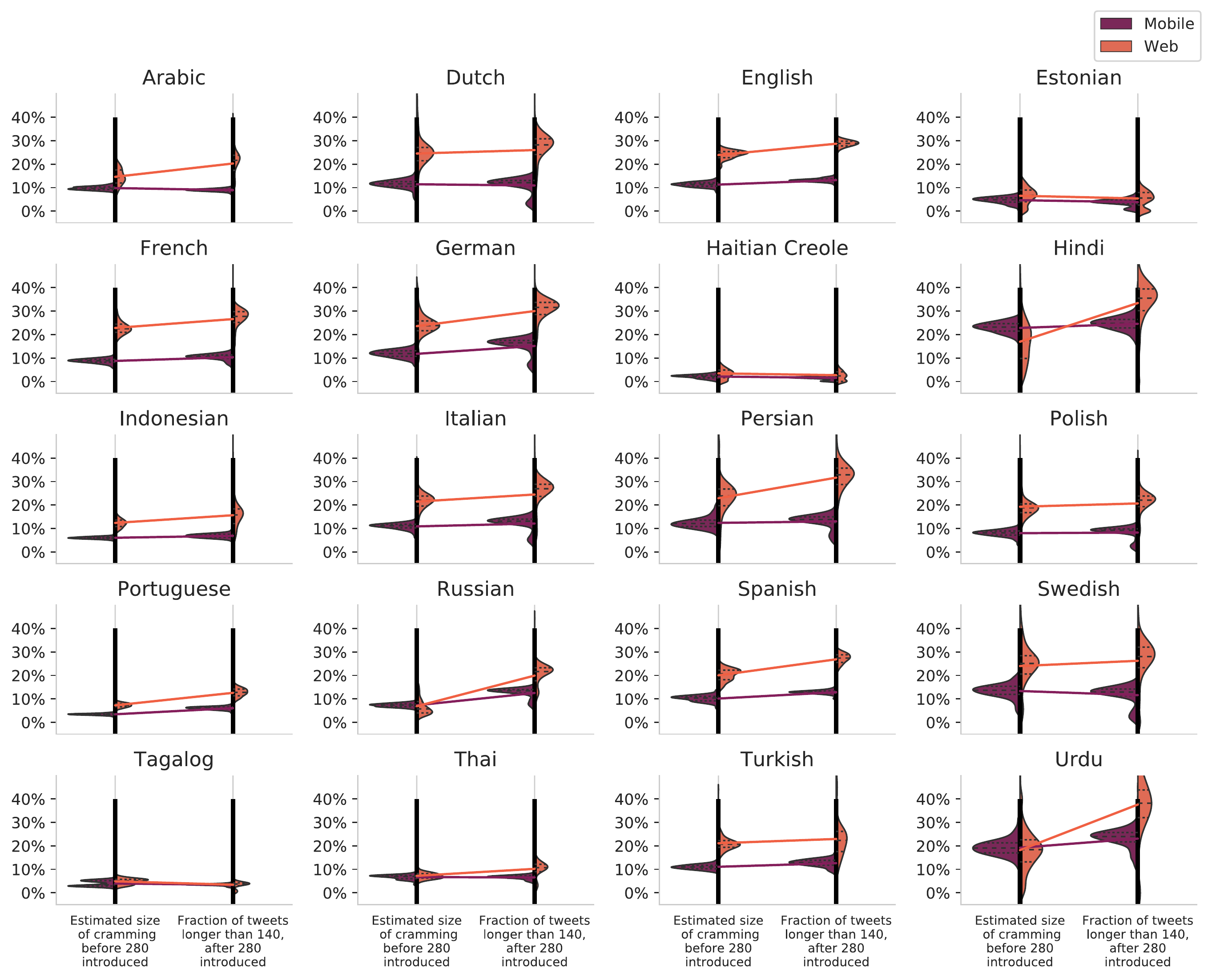}
    \caption{
    \textbf{Estimated \vs\ actual fraction of tweets impacted by the length limit, across languages.} Across 20 languages where 280 character limit was introduced, the estimated size of cramming before the intervention (on the left) and the fraction of tweets longer than 140 characters after the intervention (on the right). The point distribution across all days is displayed. Horizontal dashed lines mark the quartiles. The quantities are shown separately for mobile (in purple) and Web (in orange). Lines connect the means for mobile (in purple) and Web (in orange). Languages are sorted alphabetically.}
    \label{panel1}
\end{figure}

\begin{figure}[t]
    \centering
    \includegraphics[width= \textwidth]{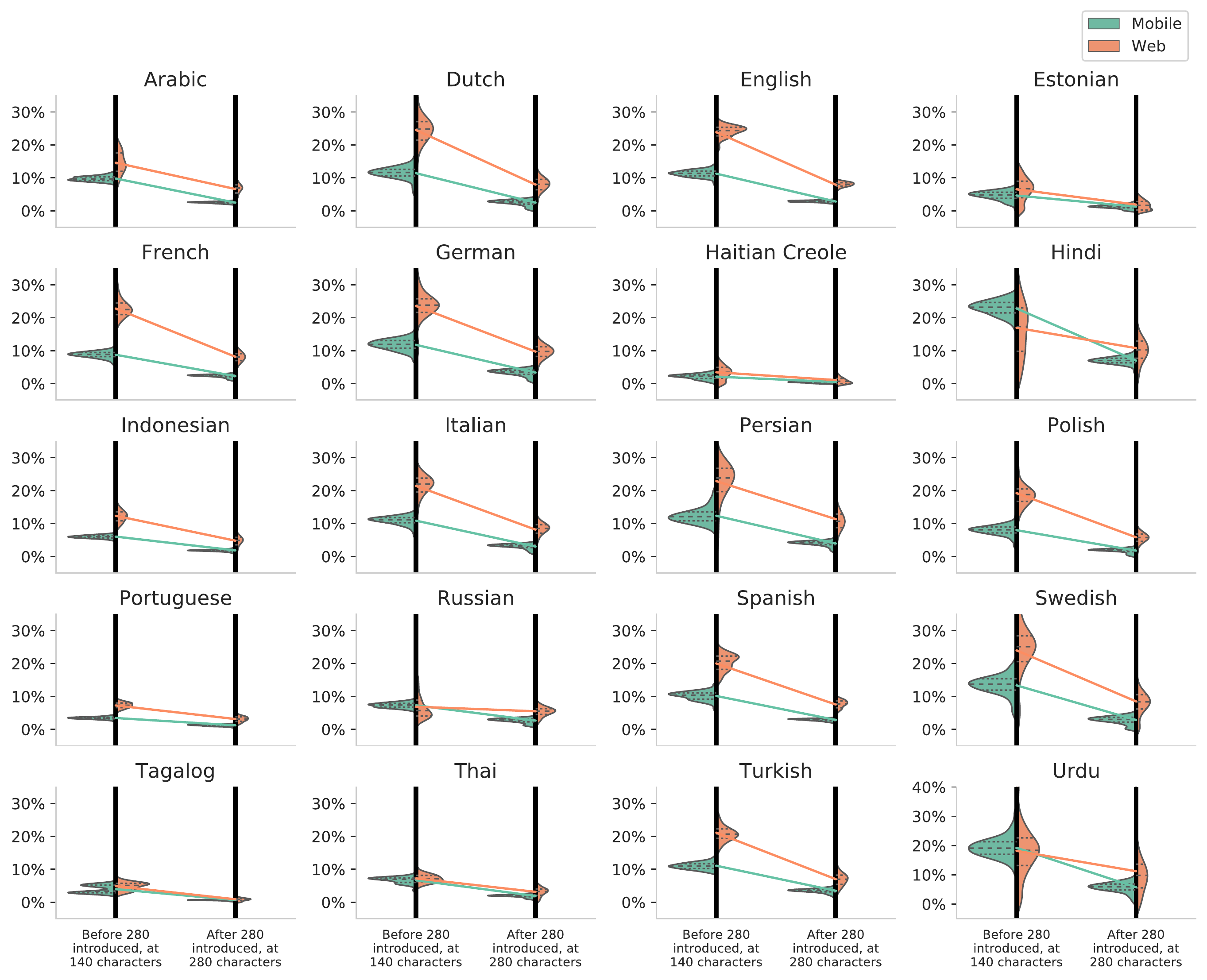}
    \caption{
    \textbf{Cramming at the enforced character limit, across languages.} Across 20 languages where 280 character limit was introduced, the estimated size of cramming before the intervention at 140 characters (on the left) and the estimated size of cramming after the intervention at 280 characters (on the right). The point distribution across all days is displayed. Horizontal dashed lines mark the quartiles. The quantities are shown separately for mobile (in green) and Web (in orange). Lines connect the means for mobile (in green) and Web (in orange). Languages are sorted alphabetically.}
    \label{panel2}
\end{figure}

Finally, for completeness, Figures \ref{panel1} and \ref{panel2} summarize the size of estimated cramming and the fraction of tweets longer than 140 characters across languages and devices.

